\RequirePackage{snapshot}

\documentclass[letterpaper,twocolumn,10pt]{article}

\usepackage{usenix2019_v3}

\usepackage{array}

\usepackage{tikz}
\usepackage{pifont}  

\usepackage[utf8]{inputenc}
\usepackage[T1]{fontenc}
\usepackage{comment}
\usepackage{dblfloatfix}

\usepackage{listings}
\usepackage{multicol}
\usepackage{xcolor}
\usepackage{textcomp}
\usepackage{caption}

\definecolor{codegreen}{rgb}{0,0.6,0}
\definecolor{codegray}{rgb}{0.5,0.5,0.5}
\definecolor{codepurple}{rgb}{0.58,0,0.82}
\definecolor{backcolour}{rgb}{0.95,0.95,0.92}

\lstdefinestyle{customc}{
  belowcaptionskip=1\baselineskip,
  breaklines=true,
  frame=single,
  xleftmargin=0.35cm,
  xrightmargin=0.15cm,
  numbers=left,
  numbersep=5pt,  
  language=C,
  showstringspaces=false,
  basicstyle=\footnotesize\ttfamily,
  keywordstyle=\bfseries\color{green!40!black},
  commentstyle=\itshape\color{purple!40!black},
  identifierstyle=\color{blue},
  stringstyle=\color{orange},
}

\lstdefinestyle{customcArianeExploit1}{
  breaklines=true,
  frame=single,
  xleftmargin=0.4cm,
  xrightmargin=0.2cm,
  numbers=left,
  numbersep=5pt,  
  language=C,
  showstringspaces=false,
  basicstyle=\footnotesize\ttfamily,
  keywordstyle=\bfseries\color{green!40!black},
  commentstyle=\itshape\color{purple!60!black},
  identifierstyle=\color{blue},
  stringstyle=\color{yellow!50!black},
  morekeywords={asm},
  keywordstyle=[2]\bfseries\color{brown!60!black},
}

\lstdefinestyle{customcArianeExploit}{
  breaklines=true,
  frame=single,
  xleftmargin=0.4cm,
  xrightmargin=0.2cm,
  numbers=left,
  numbersep=5pt,  
  language=C,
  showstringspaces=false,
  basicstyle=\footnotesize\ttfamily,
  keywordstyle=\bfseries\color{blue},
  commentstyle=\itshape\color{green!50!black},
  identifierstyle=\color{black},
  stringstyle=\color{brown},
  morekeywords={asm},
  keywordstyle=[2]\bfseries\color{black},
}

\lstdefinestyle{customlog}{
  breaklines=true,
  frame=single,
  xleftmargin=0.35cm,
  xrightmargin=0.15cm,
  numbers=left,
  numbersep=5pt,  
  language=C,
  showstringspaces=false,
  basicstyle=\footnotesize\ttfamily,
  keywordstyle=\color{blue},
  commentstyle=\itshape\color{purple!40!black},
  identifierstyle=\color{blue},
  stringstyle=\color{orange},
  keywords=[2]{INFO},
  keywords=[3]{ERROR},x
  keywordstyle=[2]\bfseries\color{green!40!black},
  keywordstyle=[3]\bfseries\color{red!500!black},
}

\definecolor{verilogcommentcolor}{RGB}{104,180,104}
\definecolor{verilogkeywordcolor}{RGB}{49,49,255}
\definecolor{verilogsystemcolor}{RGB}{128,0,255}
\definecolor{verilognumbercolor}{RGB}{255,143,102}
\definecolor{verilogstringcolor}{RGB}{160,160,160}
\definecolor{verilogdefinecolor}{RGB}{128,64,0}
\definecolor{verilogoperatorcolor}{RGB}{0,0,128}

\lstdefinestyle{prettyverilog}{
   language           = Verilog,
   commentstyle       = \color{verilogcommentcolor},
   alsoletter         = \$'0123456789\`,
   literate           = *{+}{{\verilogColorOperator{+}}}{1}%
                         {-}{{\verilogColorOperator{-}}}{1}%
                         {@}{{\verilogColorOperator{@}}}{1}%
                         {;}{{\verilogColorOperator{;}}}{1}%
                         {*}{{\verilogColorOperator{*}}}{1}%
                         {?}{{\verilogColorOperator{?}}}{1}%
                         {:}{{\verilogColorOperator{:}}}{1}%
                         {<}{{\verilogColorOperator{<}}}{1}%
                         {>}{{\verilogColorOperator{>}}}{1}%
                         {=}{{\verilogColorOperator{=}}}{1}%
                         {!}{{\verilogColorOperator{!}}}{1}%
                         {^}{{\verilogColorOperator{^}}}{1}%
                         {|}{{\verilogColorOperator{|}}}{1}%
                         {=}{{\verilogColorOperator{=}}}{1}%
                         {[}{{\verilogColorOperator{[}}}{1}%
                         {]}{{\verilogColorOperator{]}}}{1}%
                         {(}{{\verilogColorOperator{(}}}{1}%
                         {)}{{\verilogColorOperator{)}}}{1}%
                         {,}{{\verilogColorOperator{,}}}{1}%
                         {.}{{\verilogColorOperator{.}}}{1}%
                         {~}{{\verilogColorOperator{$\sim$}}}{1}%
                         {\%}{{\verilogColorOperator{\%}}}{1}%
                         {\&}{{\verilogColorOperator{\&}}}{1}%
                         {\#}{{\verilogColorOperator{\#}}}{1}%
                         {\ /\ }{{\verilogColorOperator{\ /\ }}}{3}%
                         {\ _}{\ \_}{2}%
                        ,
   morestring         = [s][\color{verilogstringcolor}]{"}{"},%
   identifierstyle    = \color{black},
   vlogdefinestyle    = \color{verilogdefinecolor},
   vlogconstantstyle  = \color{verilognumbercolor},
   vlogsystemstyle    = \color{verilogsystemcolor},
   basicstyle         = \scriptsize\fontencoding{T1}\ttfamily,
  columns=fullflexible, 
   keywordstyle       = \bfseries\color{verilogkeywordcolor},
    morekeywords      = {val, when, port, coverage},
   numbers            = left,
   numbersep          = 5pt,
   tabsize            = 2,
   escapeinside       = {/*!}{!*/},
   upquote            = true,
   sensitive          = true,
   showstringspaces   = false, 
   frame              = single,
   breaklines         = true,
   abovecaptionskip   = 0pt,
   belowcaptionskip   = 2pt,   
   xleftmargin        =0.35cm,
   xrightmargin       =0.15cm,
   captionpos         = t
}

\makeatletter

\newcommand\language@verilog{Verilog}
\expandafter\lst@NormedDef\expandafter\languageNormedDefd@verilog%
  \expandafter{\language@verilog}
  
\lst@SaveOutputDef{`'}\quotesngl@verilog
\lst@SaveOutputDef{``}\backtick@verilog
\lst@SaveOutputDef{`\$}\dollar@verilog

\newcommand\getfirstchar@verilog{}
\newcommand\getfirstchar@@verilog{}
\newcommand\firstchar@verilog{}
\def\getfirstchar@verilog#1{\getfirstchar@@verilog#1\relax}
\def\getfirstchar@@verilog#1#2\relax{\def\firstchar@verilog{#1}}

\newcommand\addedToOutput@verilog{}
\lst@AddToHook{Output}{\addedToOutput@verilog}

\newcommand\constantstyle@verilog{}
\lst@Key{vlogconstantstyle}\relax%
   {\def\constantstyle@verilog{#1}}

\newcommand\definestyle@verilog{}
\lst@Key{vlogdefinestyle}\relax%
   {\def\definestyle@verilog{#1}}

\newcommand\systemstyle@verilog{}
\lst@Key{vlogsystemstyle}\relax%
   {\def\systemstyle@verilog{#1}}

\newcount\currentchar@verilog
  
\newcommand\@ddedToOutput@verilog
{%
   \ifnum\lst@mode=\lst@Pmode%
      \expandafter\getfirstchar@verilog\expandafter{\the\lst@token}%
      \expandafter\ifx\firstchar@verilog\backtick@verilog
         \let\lst@thestyle\definestyle@verilog%
      \else
         \expandafter\ifx\firstchar@verilog\dollar@verilog
            \let\lst@thestyle\systemstyle@verilog%
         \else
            \expandafter\ifx\firstchar@verilog\quotesngl@verilog
               \let\lst@thestyle\constantstyle@verilog%
            \else
               \currentchar@verilog=48
               \loop
                  \expandafter\ifnum%
                  \expandafter`\firstchar@verilog=\currentchar@verilog%
                     \let\lst@thestyle\constantstyle@verilog%
                     \let\iterate\relax%
                  \fi
                  \advance\currentchar@verilog by \@ne%
                  \unless\ifnum\currentchar@verilog>57%
               \repeat%
            \fi
         \fi
      \fi
   \fi
}

\lst@AddToHook{PreInit}{%
  \ifx\lst@language\languageNormedDefd@verilog%
    \let\addedToOutput@verilog\@ddedToOutput@verilog%
  \fi
}

\newcommand{\verilogColorOperator}[1]
{%
  \ifnum\lst@mode=\lst@Pmode\relax%
   {\bfseries\textcolor{verilogoperatorcolor}{#1}}%
  \else
    #1%
  \fi
}

\makeatother

\lstdefinestyle{mystyle}{
    commentstyle=\textit,
    keywordstyle=\textbf,
    stringstyle=\color{codepurple},
    basicstyle=\ttfamily,
    breakatwhitespace=false,         
    breaklines=true,      
    frame=single, 
    framexleftmargin=\parindent,
    captionpos=b,                    
    keepspaces=true,                 
    numbers=left,    
    numberstyle=\normalsize,
    stepnumber=1,
    numbersep=5pt,   
    xleftmargin=1.5\parindent,
    showspaces=false,                
    showstringspaces=false,
    showtabs=false,                  
    tabsize=2
}

\usepackage{mathtools}
\usepackage{tabularx}

\usepackage{enumitem,amssymb}
\newlist{todolist}{itemize}{2}
\setlist[todolist]{label=$\square$}

\usepackage[ruled,vlined]{algorithm2e}
\usepackage{algorithmic}

\usepackage{etoolbox}
\AtBeginEnvironment{quote}{\singlespacing\small}

\usepackage{graphicx}

\usepackage{diagbox}

\usepackage{multirow}
\usepackage[symbol]{footmisc}

\newcommand*\cc[1]{\tikz[baseline=(char.base)]{
      \textcolor{black}{      \node[shape=circle,draw,inner sep=1pt] (char) {#1};}}}

\makeatletter
\def\BState{\State\hskip-\ALG@thistlm}

\newcommand{\thickhline}{%
    \noalign {\ifnum 0=`}\fi \hrule height 1pt
    \futurelet \reserved@a \@xhline
}

\makeatother

\newcommand{\greencmark}{{\color{green}\ding{51}}}%
\newcommand{\redxmark}{{\color{red}\ding{55}}}%

\usepackage[english]{babel}
\newcounter{challenge}

\usepackage[english]{babel}
\newcounter{bug}
\renewcommand*{\thebug}{\textbf{B\arabic{bug}}}

\newenvironment{bug}[1][]{\refstepcounter{bug}\par\smallskip
\noindent\textbf{Bug \thebug#1}\rmfamily}

\usepackage[english]{babel}
\newenvironment{bug_ns}[1][]{\refstepcounter{bug}
\noindent\textbf{Bug \thebug#1}\rmfamily}

\usepackage{tablefootnote}

\usepackage[normalem]{ulem}

\newcommand{\ourtool}{\emph{TheHuzz}}
\newcommand{\ariane}{Ariane}
\newcommand{\morkx}{mor1kx}

\newcommand{\difuzzrtl}{DifuzzRTL}
\newcommand{\rfuzz}{RFUZZ}
\newcommand{\rocketcore}{Rocket Core}
\newcommand{\orth}{or1200}

\usepackage{xcolor}
{}
{}
\def\todoen{0}
\def\noteen{0}
\def\removesen{0}
\def\adden{0}

\if 1\todoen 
\newcommand\todo[1]{\textcolor{red}{\textbf{TODO:} #1}}
\else
\newcommand\todo[1]{}
\fi

\if 1\noteen
\newcommand\note[1]{\textcolor{blue}{\textbf{Note:} #1}}
\else
\newcommand\note[1]{}
\fi

\if 1\removesen
\newcommand\red[1]{\textcolor{red!40!white}{\sout{#1}}}
\else
\newcommand\red[1]{}
\fi

\if 1\adden
\newcommand\blue[1]{\textcolor{blue}{#1}}
\else
\newcommand\blue[1]{{#1}}
\fi

\begin{document}

\date{}

\title{\Large \bf \ourtool{}: Instruction Fuzzing of Processors \blue{Using Golden-Reference Models} \\ for Finding Software-Exploitable Vulnerabilities}

\author{
{\rm Rahul Kande$^\dagger$, Addison Crump$^\dagger$, Garrett Persyn$^\dagger$, Patrick Jauernig$^\ast$, Ahmad-Reza Sadeghi$^\ast$,}\\ 
{\rm Akash Tyagi$^\dagger$, Jeyavijayan Rajendran$^\dagger$}\\
$^\dagger$Texas A\&M University, College Station, USA. $^\ast$Technische Universit\"at Darmstadt, Germany.\\
{\tt rahulkande@tamu.edu, addisoncrump@tamu.edu, gpersyn@tamu.edu,}\\
{\tt patrick.jauernig@trust.tu-darmstadt.de, ahmad.sadeghi@trust.tu-darmstadt.de,}\\
{\tt tyagi@tamu.edu, jv.rajendran@tamu.edu}
}

\maketitle


\begin{abstract}
The increasing complexity of modern processors poses many challenges to existing hardware verification tools and methodologies for detecting security-critical bugs. Recent
attacks on processors 
have shown the fatal consequences of
uncovering and exploiting hardware vulnerabilities.

Fuzzing has emerged as a promising technique for detecting software vulnerabilities. 
Recently, a few hardware fuzzing techniques have been proposed. However, they suffer from several limitations, including non-applicability to commonly-used Hardware Description Languages (HDLs) like Verilog and VHDL, \red{requiring}\blue{the need for} significant human intervention, and inability to capture many intrinsic hardware behaviors, such as signal transitions and floating wires.

In this paper, we present the design and implementation of a novel hardware fuzzer, \ourtool{}, that overcomes the aforementioned limitations
and significantly improves the state of the art.
We analyze the intrinsic behaviors of hardware designs in HDLs 
and then \red{formulate}\blue{measure} the coverage metrics that model such behaviors. 
\ourtool{} generates assembly-level instructions to increase the desired coverage values, 
thereby finding many hardware bugs that are exploitable from software. 
We evaluate \ourtool{} on \red{three}\blue{four} popular open-source processors and achieve 
\blue{1.98$\times$ and 3.33$\times$ the speed } compared to the industry-standard random regression approach \blue{and the state-of-the-art hardware fuzzer, DiffuzRTL, respectively}. Using \ourtool{}, we detected \red{10}\blue{11} bugs in these processors, including 8 new vulnerabilities, and we demonstrate exploits using the detected bugs. We also show that \ourtool{} overcomes the limitations of formal verification tools from the semiconductor industry by comparing its findings  \red{ to the ones from}\blue{ to those discovered by} the Cadence JasperGold tool.

\end{abstract}

\note{Use the punchline: fuzzing allows us to do more EVIL testing. We cover more and faster without compromising on coverage.}

\section{Introduction} \label{sec:intro}

Modern processors are becoming increasingly complex with sophisticated functional and security mechanisms and extensions. This development, however, increases the chance of introducing vulnerabilities into the hardware design and implementation which can lead to errors and exploitation attacks with fatal consequences. Hardware vulnerabilities range from functional bugs~(e.g., \cite{intel_fdiv_bug})
to emerging security-critical vulnerabilities that have been uncovered and exploited\red{ recently}
(e.g., \cite{kocher2019spectre},\cite{lipp2018meltdown}), and \red{not only }\blue{both }affect commodity processors \red{but also}\blue{and} 
their dedicated security extensions (e.g., \cite{SgxPectre}, \cite{foreshadow}).
The hardware common weakness enumeration (CWE) lists numerous hardware vulnerabilities 
whose impact spans not only the hardware but also software~\cite{hardware_cwe}. 
It is crucial to discover hardware vulnerabilities in the early stages of the design cycle.

Various hardware vulnerability detection techniques and tools have been proposed or developed by both academia and industry, such as formal verification~\cite{cadence_home,synopsys_home, modelsim, solidify, onespin_home, wile2005comprehensive,dessouky2019hardfails,clarke2011model,rajendran2015detecting}, run-time detection~\cite{hicks2015specs,sarangi2006phoenix,wagner2007engineering}, information flow tracking \cite{glift11,clepsydra17,caisson11,sapper14,secverilog15}, and the recent efforts towards fuzzing hardware \cite{hyperfuzzing,google_fuzzer,  laeufer2018rfuzz, difuzz}, which is the focus of this paper.

While formal verification tools can efficiently find  bugs in smaller designs, they are unable to cope with the increasing complexity of modern, large designs and are becoming less efficient in detecting bugs,  especially in detecting security vulnerabilities \cite{dessouky2019hardfails,marshall2019hardware,yang2003current,chen2017challenges,farahmandi2019system}. One particular reason is that these tools rely heavily on human expertise to engineer\red{/}\blue{ or }specify ``attack scenarios'' for verification.
For instance, the popular industrial formal verification tool, Cadence's JasperGold~\cite{cadence_home}, has been evaluated against a crowd-sourced vulnerability detection effort from 54 competing teams participating in a hardware capture-the-flag competition~\cite{dessouky2019hardfails}. The results were based on security bugs mimicking real-world common vulnerabilities and exposures (CVEs)~\cite{hp-laserjet-cve,microsoft-hypervisor-cve,apple-audiocodecs-cve,dell-bios-cve,google-mediatek-cve,samsung-pagetable-cve}. While JasperGold \red{could not detect 52\%}\blue{detected 48\%} of the bugs, manual inspection with simulation \red{only failed to detect 39\%}\blue{detected 61\%} of the bugs, highlighting
issues like state explosion and scalability of the existing techniques, amongst others.

Another approach to find hardware security  bugs is run-time detection techniques, which hardcode assertions in hardware to check security violations at runtime~\cite{hicks2015specs,sarangi2006phoenix,wagner2007engineering}. 
However, these techniques detect bugs only post-fabrication\red{;} \blue{and} unlike software, hardware is not easily patchable. 

Information-flow tracking (IFT) techniques analyze the target hardware to detect security vulnerabilities by labeling all the input signals and propagating this label throughout the design to identify information leakage or tampering of critical data~\cite{glift11,clepsydra17,caisson11,sapper14,secverilog15}. 
Although IFT can analyze designs with several thousand lines of code, the labels often get polluted with unwanted signals, resulting in a high number of false positives. 
The initial labels have to be assigned manually, which can be error-prone, and  require expert knowledge of the design. 

Hence, there is an increasing need for methodologies and tools to detect hardware vulnerabilities that are scalable to large and complex designs, highly automatic, effective and efficient in detecting security-critical vulnerabilities that are exploitable (and not just only functional bugs), compatible with existing chip design and verification flows, applicable to different hardware models (register transfer-level, gate-level, transistor-level, taped-out chip), and account for different hardware behaviors (signal transitions, finite state machines, and floating wires). 

\red{In this context, a}\blue{A} promising technique extensively used for software vulnerability detection is fuzzing.
Fuzzing uses random generation of test cases to detect invalid states in the target\cite{fuzzing-survey}.
While it seems natural to apply or extend a software fuzzer to detect security bugs in hardware~\cite{google_fuzzer,hyperfuzzing}, such approaches do not capture hardware-intrinsic behaviors, for instance, signal transitions of wires, finite-state machines (FSMs), and floating wires, defined in hardware description languages (HDLs) like Verilog and VHDL.
We will discuss these challenges in Section~\ref{sec:challenges}.

So far, there have been a few proposals towards fuzzing hardware~\cite{laeufer2018rfuzz},~\cite{hyperfuzzing},~\cite{google_fuzzer}, and~\cite{difuzz}.
However, as we elaborate in detail in Section~\ref{sec:related_work}, these proposals suffer from various limitations such as lack of support for commonly-used HDLs such as VHDL and Verilog ~\cite{laeufer2018rfuzz} \blue{or only partially supporting their constructs~\cite{google_fuzzer}}, strong reliance on human intervention~\cite{hyperfuzzing}, 
\red{
requirement of equivalency checks between the hardware model and the software model derived from the hardware~\cite{google_fuzzer}
}
and the inherent inability of capturing many hardware behaviors, including transitioning of logical values in wires and of floating wires~\cite{difuzz}. 

\noindent\\
\textbf{Our goals and contributions.}
We present the design and implementation of a novel hardware fuzzer, \ourtool{}. It tackles the challenges of building a hardware fuzzer (cf. Section \ref{sec:challenges}) and  addresses the aforementioned shortcomings of the current hardware fuzzing proposals (cf. Section \ref{sec:our_hw_fuzzer}).
We analyze the intrinsic behaviors of hardware designs and describe appropriate coverage metrics of the HDL to capture such behaviors. Given the importance of software-exploitable hardware vulnerabilities~\cite{dessouky2019hardfails,marshall2019hardware,yang2003current,chen2017challenges,farahmandi2019system}, 
\ourtool{} fuzzes the target hardware by testing instruction sequences, thereby discovering security bugs that are exploitable by the software code which executes such instruction sequences. 
Through a built-in optimizer, \ourtool{} can select the best instructions and mutation techniques to achieve the best coverage. 
  
\ourtool{} (i)~supports commonly-used HDLs like Verilog and VHDL,
(ii)~is compatible with conventional \blue{industry-standard} IC design and verification flow, 
(iii)~detects software-exploitable hardware vulnerabilities,
(iv)~accounts for different hardware behaviors,
(v)~does not require knowledge of the design,
(vi)~is scalable to large-scale designs,
and (vii)~does not need human intervention.

In summary, our main contributions are:

\begin{itemize}[align=parleft,leftmargin=*]
\item We present a novel hardware fuzzer, \ourtool{}, (Section~ \ref{sec:our_hw_fuzzer}), which uses coverage metrics that capture a wide variety of hardware behaviors, including signal transitions, floating wires, multiplexers, along with combinational and sequential logic. 
\ourtool{} optimizes the selection of the best instructions and mutation techniques  and can achieve high coverage rates (c.f. Section \ref{sec:our_fuzzing_tool}). 
\blue{Our fuzzer achieves 1.98$\times$ and 3.33$\times$ the speed compared to the industry-standard random regression approach and the state-of-the-art hardware fuzzer, DiffuzRTL, respectively} (cf. Section~\ref{sec:coverageAnalysis}).

\item We extensively evaluate our fuzzer, \ourtool{}, on \red{three}\blue{four} well-known and complex real-world open-source processor designs from two different open-source instruction set architectures (ISAs): (i)~or1200 processor (OpenRISC ISA), (ii)~mor1kx processor (OpenRISC ISA), \red{and} (iii)~Ariane processor (RISC-V ISA)\blue{, and (iv)~\rocketcore{}}. All these processors can run Linux-based operating systems and are used in multiple hardware verification research studies~\cite{dessouky2019hardfails,hicks2015specs,zhang2017scifinder,zhang2018end}.
\item \ourtool{} found  \blue{11} bugs that are software exploitable in three different processors; eight of them are new bugs. 
 We also showcase two attacks from unprivileged software exploiting vulnerabilities found by \ourtool{}~(cf. Section~\ref{sec:bugs}).

\item We perform an investigation of the bugs detected by \ourtool{}
using a leading formal verification tool, Cadence's JasperGold~\cite{jasperGold} (cf. Section~\ref{sec:formalComparison}).
\ourtool{} overcomes the limitations of JasperGold:
state explosion, \red{resource-intensive}\blue{intensive resource consumption}, reliance upon \blue{error-prone} human expertise\red{ and thus, error-prone}, and \red{requiring }\blue{a requirement of }prior knowledge of hardware vulnerabilities or security properties.
 
\item To foster research in the area of hardware fuzzing, we plan to open-source the code of \ourtool{} to provide the community a framework to build upon.

\end{itemize}

\section{Background}
\label{sec:background}

The growing number of attacks that exploit hardware vulnerabilities from software \cite{intel_fdiv_bug,kocher2019spectre,lipp2018meltdown,rowhammer_0,rowhammer_1,rowhammer_2,rowhammer_3,clkscrew2017tang,voltjockey,kenjar2020v0ltpwn,SgxPectre,foreshadow} calls for new and effective vulnerability detection techniques in hardware that address the limitations of existing methods and tools, such as state-space explosion, modeling hardware-software interactions, and the need for manual analysis.

\subsection{Fuzzing}
Fuzzing techniques are shown to be highly effective in detecting software vulnerabilities~\cite{takanen2018fuzzing,citeafl,fuzzing-survey,ossfuzz,fuzzilli,webappfuzz,syzkaller,freedom}. Fuzzing generates test inputs and 
simulates the target design to detect vulnerabilities in it. The inputs are generated by \textit{mutating} the previous inputs, which are generated from seeds. Mutation techniques modify the input by performing predefined operations, including bit-flip, clone, and swap. 
The mutation process also generates invalid inputs, testing the design outside the specification. 
In the past, fuzzers were created specifically to target different kinds of software: binary targets\cite{citeafl}, JIT compilers\cite{fuzzilli}, web applications\cite{webappfuzz}, and operating systems\cite{syzkaller}. 
Thus, specialized fuzzers conform to the needs of each target type.
Fuzzers have seen use from both independent researchers and organizations as an additional verification step, most notably that of Google's OSS Fuzz~\cite{ossfuzz}, which actively fuzzes a plethora of software on their ClusterFuzz platform~\cite{clusterfuzz}. 
Fuzzers are highly successful in detecting software vulnerabilities 
as they are automated, are scalable to large codebases, do not require the knowledge of the underlying system, and are highly efficient in detecting many security vulnerabilities.

Unfortunately, comparable approaches for hardware fuzzing are still in their infancy. Hardware-specific behaviors pose several challenges to the design of hardware fuzzers, which we  present in this section. 
However, before we consider the natural question of why one cannot trivially adopt the advances of software fuzzers
for hardware, 
we briefly explain the typical hardware (security) development life cycle.

\subsection{Hardware Development Lifecycle}\label{sec:HSDL}

The hardware development lifecycle~\cite{sandia_sdl, badawy2012system, vasudevan2006introduction, molina2007functional} typically begins with 
a design exploration driven by the market segment served by the product. 
Architects then engineer the optimal architecture while trading off among performance, area, and power, and the associated microarchitectural features. 
Designers implement all the microarchitectural modules using Hardware Description Languages (HDLs), which are usually written at the Register-Transfer Level (RTL). 
To this end, popular HDLs like Verilog and VHDL are used to describe complex hardware structures such as buses, controllers as finite state machines (FSMs), queues, and datapath units like adders, multipliers, etc. 
Electronic design automation (EDA) tools synthesize the RTL models into gate-level designs, which realize the hardware using Boolean gates, multiplexers, wires, and state elements like flip flops.
EDA tools then synthesize the gate-level design into transistor-level and eventually to layout, which is then sent to the foundry for manufacturing.

Most of the design effort and time \blue{spent by designers} goes into \red{designers }manually writing HDLs at the RTL as the rest of the steps are highly automated. Unfortunately, writing HDL at the RTL is error-prone~\cite{badawy2012system, vasudevan2006introduction, molina2007functional}.
Thus, the verification team checks if the design at its various stages meets the required specification or not using functional, formal, and simulation-based tools; if the design does not meet the specification, the designers patch the bugs, and the process is repeated until the design passes the verification tests. 
To this end, companies typically develop a golden reference model (GRM)\footnote{The GRM for hardware is similar to a test oracle in software which helps verify the result of a program's execution~\cite{testoracle}.
} for industry designs to be used with the conventional verification flows. 
 GRMs are often written at a higher abstraction level (e.g., for RTL, the GRM is a software model of the hardware). 
 Verification techniques usually compare the outputs of RTL and the GRM to find any mismatches, which will reveal the bugs. 
 The accuracy of these techniques is further increased by comparing not only
 the final outputs but also 
 the values of intermediate registers 
 and by performing comparisons after every clock cycle. 
They perform similar tests on the gate-level design and the fabricated chip; for these models, the  adjacent abstraction level acts as the GRM. 
Similarly, post-manufacturing, testing of the fabricated chips is performed to weed out the faulty chips.

When the architecture of the chip is designed, the security team concurrently identifies the threat model, security features, and assets. 
During the design phase, the security team performs security testing, starting with the RTL model via simulation and emulation, formal verification, and manual review of RTL codes. 
Post-deployment, the security engineers provide support and patch any bugs, if possible.

\section{Challenges of Hardware Fuzzing }
\label{sec:challenges}
In this section, we outline the challenges that arise when analyzing hardware using fuzzing.  We first elaborate on the problems that one encounters when deploying existing software fuzzers to analyze hardware.
Then we discuss challenges that need to be tackled when designing and implementing a dedicated hardware fuzzer.

\subsection{Fuzzing Hardware with Software Fuzzers}\label{sec: fuzzing_hw_w_sw}

There are two ways to fuzz hardware with software fuzzers: (i)~using software fuzzers directly on the hardware, and (ii)~fuzzing hardware as software. However, both approaches face several limitations.
\noindent \\
{\textbf{Problems with using software fuzzers directly on hardware.}}
First, software fuzzers rely on a different behavior in vulnerability detection. They rely on software abstractions to find a bug by using the operating system or instrumenting software to monitor failure detection~\cite{citevalgrind,citeasan}. 
\red{Unlike software, hardware does not \textit{crash}, indicating the bug detection used in most software fuzzers}
\blue{Most software fuzzers use crashes to detect bugs, but hardware does not \textit{crash}}~\cite{google_fuzzer}. Thus, hardware fuzzers need to find their  equivalent of crashes and memory leaks. 
Second, hardware simulations are slow. 
Typically, given a function, executing its software equivalent is faster than simulating its hardware model. 
Parallelization of hardware simulation is difficult due to the complex interdependencies in the hardware design~\cite{chen2017challenges}. 
Third, many software fuzzers rely on {\it instrumentation} of the software program to obtain feedback (e.g., AFL's \cite{citeafl}) and use custom compilers (e.g., \textit{afl-gcc}) to instrument the code \cite{hsu2018instrim,gan2018collafl,li2017steelix,citeafl}, but these compilers will not be able to instrument the hardware designs since they do not support HDLs  such as Verilog and VHDL. 
\blue{One of the prior works, RFUZZ~\cite{laeufer2018rfuzz} made the first attempt towards solving this challenge: it uses hardware simulators to compile the hardware and applies a modified version of software fuzzer, AFL~\cite{citeafl} to fuzz the hardware. However, this fuzzer is limited in terms of the scalability~\cite{difuzz} and coverage  (cf. Appendix~\ref{App:other_cov_metrics}).}
\noindent\\
{\textbf{Problems with fuzzing hardware as software.}} \label{sec:fuzzing_in_software}
Another strategy of fuzzing hardware using software fuzzers is to convert an HDL model into an equivalent software model using tools like Verilator~\cite{verilator}, and then apply software fuzzers to the resultant software code~\cite{google_fuzzer}.
Unfortunately, converting hardware into software models poses its own set of challenges.
    
   First, applying existing software fuzzers on software models of hardware designs is, in general, inefficient.
   The software
models of hardware designs need to account for properties unique to
the working of the hardware, like computing all the register
values for every clock cycle and bit manipulation operations,
and components such as controllers, system bus, and
queues---which makes the model computationally expensive.
   Moreover, software fuzzers use program crashes and instrumented memory safety checks to detect bugs in an application; these concepts \blue{such as crashing and memory leakage detection mechanisms cannot be trivially applied to hardware~\cite{google_fuzzer}}.
   Instead, a well-defined specification to compare against is needed to detect incorrect logic implementations, timing violations, and unintended data flow or control flow.

    Second, inferring actual hardware coverage from the generated software model is difficult.
    While software and hardware line and edge/block coverage are comparable in some instances~\cite{google_fuzzer}, other forms of coverage may not be.
    A relatively simple operation in an HDL, such as bit manipulation, may be significantly more complex in software.
    Conversely, a more complex component of a hardware design, such as a multiplexer, could be represented by a simple switch statement in software.
\red{    Similarly, the software line, branch, and especially, path coverage cannot be compared to those of the hardware designs.}
    Compounding this problem further, different software modeling programs will perform the conversion differently. 
    \blue{Thus, one has to account for the effects of conversion.}
    \red{any attempt to translate or correlate software coverage metrics with those of the hardware will be specific to the software model.}

   \red{ Third, there is a {\it lack of tools} to convert hardware into equivalent software models since not many industry-standard tools---especially from large companies, e.g., Cadence, Synopsys, Mentor Graphics---exist for this purpose except for the Verilator tool~\cite {verilator}. 

    This severely limits the usability of fuzzers to a small range of hardware designs. 
    }
    
    \red{Fourth}\blue{Third}, the hardware community has developed its own standards, processes, and flows for using verification methodologies and tools over several decades of research~\cite{molina2007functional,badawy2012system,vasudevan2006introduction}.
    \blue{Any new approach has to be} \red{Converting hardware to software models is rather a new approach. It is in} \blue{compatible with the hardware verification flow, as these methodologies have specialized data structures and algorithms geared towards hardware models and behaviors.} \red{, which may not readily translate to their software equivalents. This will inevitably require revamping the entire IC verification process, which already occupies 70\% of the design effort.} 

\blue{An open-source approach to solve the many  challenges of fuzzing the software model of hardware is performed in~\cite{google_fuzzer}.
This technique derives equivalences between the coverage metrics (e.g., line and FSM) used in hardware to that of software (e.g., line and edge). 
While this approach is promising, it does not scale to complex designs such as processors, which is the focus of this work (cf. Section~\ref{sec:related_work}).}

\red{    Finally, the process of translating hardware into ``equivalent'' software models brings its own potential inaccuracies and risks. There is no practical way of telling if the software model is an accurate representation of the hardware.}

\red{    Thus, the approach of running software fuzzers to catch hardware bugs is only a first ad-hoc step in addressing this complex problem.
    In the long term, fuzzers dedicated and developed specifically for hardware are needed to meet the growing needs of hardware verification.} 

\subsection{\red{Challenges in }Creating a Hardware Fuzzer}
\label{sec:fuzzing_challenges}
 
A hardware fuzzer needs to take into account the nature and requirements of hardware to improve efficiency. For example, Syzkaller~\cite{syzkaller}, which specializes in kernel-fuzzing, incorporates system call signatures to generate better test cases. 
A hardware design fundamentally differs from any software program in terms of inputs, language used, feedback information available, and design complexity. Also, designing a hardware fuzzer has its own set of unique challenges, which are presented below.
\blue{Multiple attempts have been made in the recent past towards building hardware fuzzers~\cite{laeufer2018rfuzz, difuzz, hyperfuzzing, google_fuzzer} where each of these challenges are approached differently.}

\noindent\textbf{Input generation.}
For a hardware fuzzer to be efficient and effective, it should generate inputs in the format expected by the target processor.
Directly applying the input-generation techniques used in software fuzzing is impossible as the input formats differ: while many software fuzzers take input files or a set of values assigned to a variable, the input to hardware is mostly continuous without a defined length~\cite{google_fuzzer}.
Further, inputs to hardware can be generated at various hardware abstraction levels: architecture level, register-transfer level (RTL), gate level, and transistor level. Each level also has its own input representation, ranging from transaction packets, over continuous-time digital signals, to continuous-time analog signals.
Hence, the major challenges in input generation are to \textit{determine the suitable abstraction level to fuzz and the input representation} that maximizes the efficiency in finding vulnerabilities~\cite{chen2017challenges,molina2007functional, hyperfuzzing, laeufer2018rfuzz, difuzz, google_fuzzer} .

Another important aspect is the continuous nature of the hardware since it changes its state with every input (and/or time).
\blue{Also, multiple FSMs can run in parallel, and one or more of them could enter in deadlock states, preventing the hardware from receiving inputs from the fuzzer~\cite{chen2017challenges}.}
For instance, a password checking module could be designed to lock itself forever after one incorrect password entry unless the system is reset.
Hence, another crucial challenge is to \textit{identify situations where the hardware simulation should be stopped or reset before applying new inputs.} 

Finally, similar to how software fuzzers like syzkaller~\cite{syzkaller} encode functional dependencies (e.g., of system calls), hardware modules often need to be initialized to enable the fuzzer to test further functionality, e.g., an AES encryption module needs to be initialized with the key size and encryption mode before testing the actual encryption with plaintext and key.
\blue{\textit{Inferring these functional dependencies is highly challenging, 
} 
as such information is usually only available with a well-defined formal specification~\cite{hyperfuzzing, google_fuzzer}.}

\noindent\textbf{Feedback mechanism.} 
Exploring complex targets, especially hardware, often forces fuzzers to generate tremendous amounts of inputs, while making decisions like which mutation technique to use, when to stop mutating an input, and how to generate the seed inputs repeatedly. Rather than relying on randomly-generated inputs alone, a more efficient way is to analyze the impact of these parameters on the target processor and adapt input generation accordingly as done in feedback-guided fuzzing~\cite{kafl,li2018fuzzing}. 
\blue {Prior works~\cite{laeufer2018rfuzz,difuzz} addressed this challenge using hardware-friendly coverage metrics but fail to capture many hardware behaviors (cf. Appendix~\ref{App:other_cov_metrics}).}

\textit{Adapting software feedback mechanisms to hardware is difficult} due to the differences in execution/simulation for software and hardware~\cite{laeufer2018rfuzz, difuzz, hyperfuzzing, google_fuzzer}.
Instrumentation needs to be added to the hardware design such that the activities of different combinational and sequential structures, which are critical to the functionality of the hardware, can be traced. \red{For instance, components like registers, multiplexers, decoders, FSM controllers, shifters, arithmetic and logical units, and data buses need to be instrumented.}
Although feedback-guided fuzzers have more potential to explore complex targets, capturing, analyzing, and processing the feedback data is challenging~\cite{kafl,li2018fuzzing}. This issue will become more profound in hardware since hardware designs are slower to simulate. 
\blue{One way to speedup hardware fuzzing is to use FPGA emulation, but instrumenting a design on an FPGA is challenging~\cite{laeufer2018rfuzz,difuzz}}. 
Hence, the \textit{feedback mechanism needs to capture the complex characteristics of hardware. \red{ while processing and storing the analysis data efficiently.}}

Lastly, the performance of a fuzzer needs to be evaluated on hardware designs comparable to what is used in practice. 
However, unlike with software, commercial hardware designs like Intel's x86 processors do not have their source code available.
Hence, a key challenge is to find \textit{openly-available designs that are reasonably modern and complex. 
}
\section{Design of Our Fuzzer, \ourtool{}} \label{sec:our_hw_fuzzer} 

\ourtool{} is a novel hardware fuzzer that overcomes the challenges identified in Section~\ref{sec:fuzzing_challenges}. We directly fuzz the hardware model instead of the software model, thereby eliminating the need for hardware-to-software conversions and the associated equivalency checks. 
To overcome the slowness of hardware simulation, \ourtool{} selects the optimal instructions and mutation techniques to use. 
\ourtool{} is easily integratable with existing hardware design and verification  methodologies---thereby, easily adaptable by companies---as our approach does not require any modification to the target processor and utilizes existing hardware simulation tools and techniques.
We refer to the target processor as the design under test (DUT). \red{, in accordance with hardware verification terminology.}
Our fuzzer generates instructions as inputs to the DUT since we focus on software-exploitable processor vulnerabilities. \red{ in this work.}

\begin{figure}[ht]
    \centering
    \hspace*{0cm}
    \includegraphics[trim=24 21 20 19,clip,width=\columnwidth]{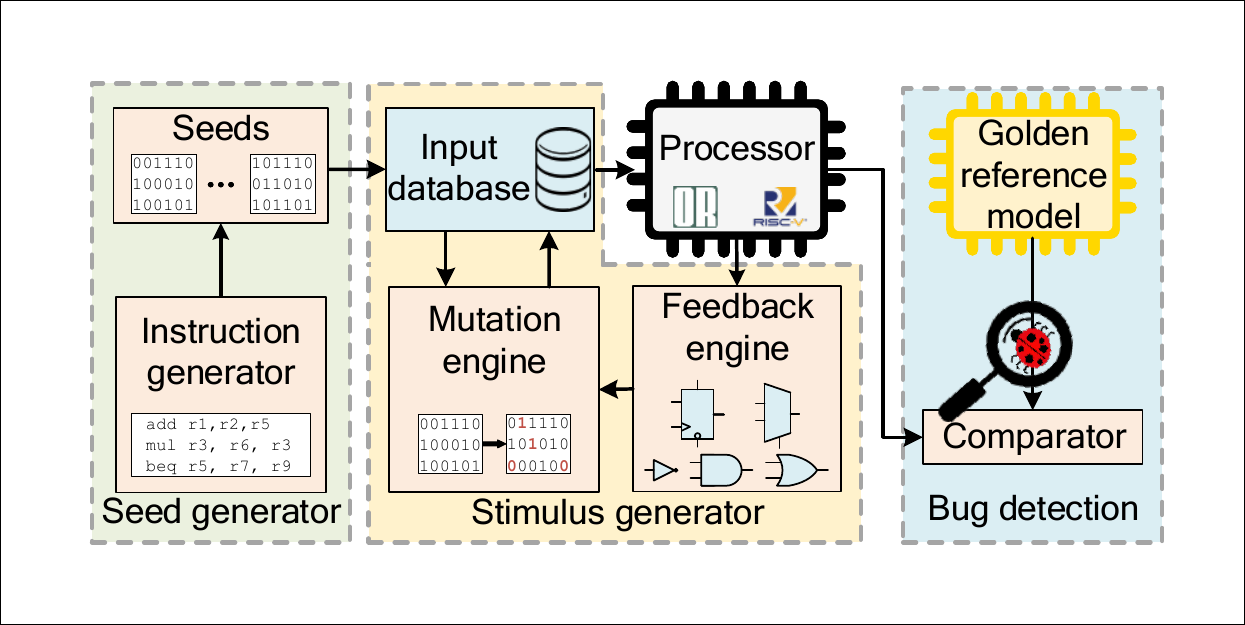}
    \caption[Hardware Fuzzer Design]{
    Framework of \ourtool{}.  
    }
    \label{fig:fuzz_des}
\end{figure}

\ourtool{} comprises three modules, as shown in Figure~\ref{fig:fuzz_des}.
First, the \textit{seed generator} starts the fuzzing process by generating an initial sequence of instructions (\textit{seeds} or \textit{seed inputs}). 
Then, the \textit{stimulus generator} generates new instruction sequences by mutating them, beginning with the \textit{seeds}.
These inputs are passed to the simulated RTL design of the DUT, which returns coverage feedback to the \textit{stimulus generator} and trace information for \textit{bug detection}.
Finally, the \textit{bug detection} mechanism compares the RTL simulation trace with that of a Golden Reference Model (GRM) to find differences in execution, and hence, find bugs.

In the following, before we explain the modules of \ourtool{}, we first analyze the intrinsic behaviors of designs at the RTL, as \ourtool{} targets such behaviors,
and describe the coverage metrics that  capture those behaviors.
Then, we describe the seed generator and stimulus generator of our fuzzer in detail and how they interact.
Finally, we detail how we optimize the mutation engine and how the bugs are detected.

\subsection{Hardware Design and Coverage Metrics}
\label{sec:hw_design}

Hardware designs at RTL consist of combinational and sequential logic. 
Combinational logic is a time-independent circuit consisting of boolean logic gates (e.g., AND, OR, XOR) and wires connecting them. 
Apart from building datpath units like adders and multipliers, these logic gates are used to build basic combinational structures like multiplexers (MUXes), demultiplexers, encoders, and decoders, which are in turn used in building complex blocks. 
\red{Sequential logic generates output based on inputs and the current state. }
Apart from combinational gates, sequential logic also uses registers, which are usually implemented using D flip-flops (DFFs). 
In the following, we explain the effectiveness of our fuzzer in capturing hardware behavior over existing hardware fuzzers using a \blue{case study}.

\let\othelstnumber=\thelstnumber
\def\createlinenumber#1#2{
    \edef\thelstnumber{%
        \unexpanded{%
            \ifnum#1=\value{lstnumber}\relax
              #2%
            \else}%
        \expandafter\unexpanded\expandafter{\thelstnumber\othelstnumber\fi}%
    }
    \ifx\othelstnumber=\relax\else
      \let\othelstnumber\relax
    \fi
}

\begin{lstlisting}[language=Verilog, label = {listing:ExampleDesChisel}, caption={\blue{Chisel code of the case study}.},style=prettyverilog,float,belowskip=-15pt,aboveskip=-5pt,firstnumber=36]
    // combinational logic for vld register
    vld := debug_en | (flush | en) // Bug b2
    
    // select signal for mux
    val sel1 = Wire(Bool())
    sel1 := ((pass === ipass) | debug_en) // Bug b1
                                
    // flush logic
    val state_f = Wire(UInt(3.W))
    when (flush & en){
        state_f := FLUSH
    }. otherwise {
        state_f := state
    }
    
    state := (!sel1 & state_f) | (sel1 & D_READ)
\end{lstlisting}

\noindent \blue{\textbf{Case study.}}
\blue{We now present a case study using a design with two bugs inspired by CVEs.
First, we explain the intended behavior and then the bugs. Then, we detail \ourtool{}'s coverage metrics and describe how they detect these bugs.}

\blue{
Consider a cache controller module---similar to the instruction cache controller of the \ariane{} processor~\cite{ariane_processor}---shown in Listing~\ref{listing:ExampleDesChisel}.
As shown in Figure~\ref{fig:ExampleDesignFSM}, the {\tt D\_READ} and the {\tt FLUSH} states determine the read operation during the debug mode and the flush operation during the normal mode, respectively, as listed in Lines 39--51\footnote{For succinctness, we ignore the other states of the cache controller.}. 
The controller enters the {\tt FLUSH} state when there is a flush command and if  the cache is enabled. 
The intended behavior of the FSM is that the read operations in the debug mode are permitted only if the user has inputted the correct password (Line 41). 
This protection mechanism allows only authorized users to read the cache in the debug mode.
The cache controller sets the valid signal ({\tt vld}) based on the flush and debug requests issued to the controller (Line 37 of ~\ref{listing:ExampleDesChisel}). 
}

\begin{figure}[tb!]
    \centering
    \hspace{-1cm}
    \includegraphics[trim=20 10 20 10,clip,width=1\columnwidth]{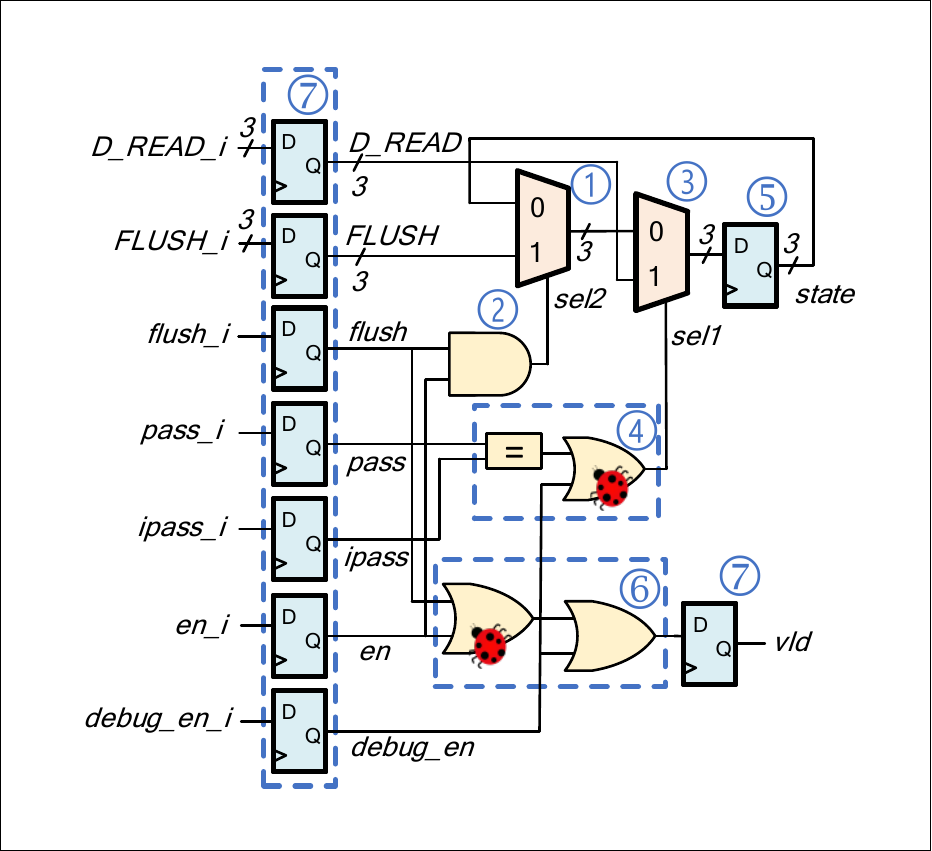}
    \caption[]{Hardware design for Listing~\ref{listing:ExampleDesChisel}.
    }
    \label{fig:ExampleDesign}
    \vspace{-.35cm}
\end{figure}

\begin{figure}[tb!]
    \centering
    \hspace{-1cm}
    \includegraphics[trim=20 20 20 20,clip,width=1\columnwidth]{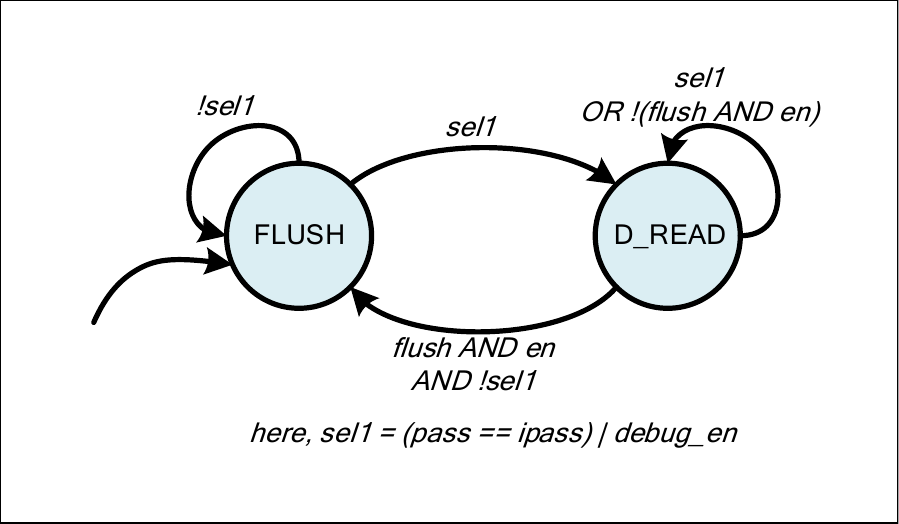}
    \caption[]{\blue{Finite state machine (FSM) of the  design in Figure~\ref{fig:ExampleDesign}.
    }}
    \label{fig:ExampleDesignFSM}
\end{figure}

\blue{
The EDA tools synthesize this RTL code into an equivalent gate-level design shown in Figure~\ref{fig:ExampleDesign}.
The MUXes \cc{1} and \cc{3} select the next state.
The combinational logic \cc{2} and \cc{4} controls the state transitions.  
The DFFs in \cc{5} hold the current state.
The EDA synthesis tools implement Line 37 as combinational logic \cc{6}.
The DFFs in \cc{7} register the inputs and outputs.
}

\blue{This design has two bugs: \textit{b1} and \textit{b2}. 
Bug \textit{b1} (Line 41 in Listing~\ref{listing:ExampleDesChisel}) is from HardFails~\cite{dessouky2019hardfails}, which  has been used for the Hack@DAC competitions, and is similar to CVE-2017-18293. 
This bug is in the combinational logic \cc{4}, where the debug read operation is access-protected but the bug allows one to perform the debug read operation illegally. 
This compromises the security of the read operations 
as it allows users without the correct password to read the cache. 
Bug \textit{b2} is similar to CVE-2019-19602 and is in the combinational logic \cc{6} that drives the \texttt{vld} register  (Line 37), allowing one to flush the cache even when it is not enabled.  
}


\blue{In \cc{1}, all the inputs of the MUX\red{along with} 
and their corresponding values on the select lines must be tested for correctness. 
For this purpose, we use {\bf branch coverage}, which tests each branching construct (the {\tt when} block of Line 45) for both ``when'' and ``otherwise'' conditions.}

\blue{In \cc{2}, 
one should check that every input combination produces the correct output value.
To this end, we use {\bf condition coverage}, which requires the condition block (i.e., the condition {\tt (flush\ \&\ en)} in the {\tt when} block of Line 45) to be tested for all possible input values and not only a subset of values.}

\blue{In \cc{5}, the value of the 3-bit register can be one of the eight possible values.
We use {\bf FSM coverage} of the {\tt state} register to check for all the eight values. This coverage captures the different FSM states and also their transitions}. 

In \blue{\cc{6}, all the input signals generating {\tt vld} should be tested for all possible values, similar to \cc{2}.
We use  {\bf expression coverage} for this purpose which requires the combinational block (i.e., the expression {\tt debug\_en|(flush|en)} in Line 37) to be tested for all possible input values.
Furthermore, expression coverage covers the select line of MUX \cc{3} and the combinational logic \cc{4} that drives it as they are defined using an expression in Line 51, unlike MUX \cc{1} which is defined as branch in lines 45--49. }

\blue{In \cc{5} and \cc{7}, the value of each D-flip flop (DFF) can be {\tt 1},  {\tt 0}, or  floating\footnote{\blue{Referred to as a high-impedance state or tristate and denoted as {\it z}. Such floating wire-related bugs (CWE-1189) have compromised systems ~\cite{dessouky2019hardfails}.}}.
We use {\bf toggle coverage} of these DFFs to check for toggling of their values among these three possibilities. 
Unlike FSM coverage, toggle coverage covers all the DFFs in the design. }
In addition, we also use {\bf statement coverage} to ensure every line of the RTL code is executed during simulation.  

\blue{\ourtool{} uses commercial industrial-standard tools---Synopsys~\cite{synopsys_vcs}, ModelSim~\cite{modelsim}, Cadence~\cite{cadence_home}---to generate the software model of hardware and extract these coverage values. 
The semiconductor industry has been using these tools for the last few decades, and its verification flow is built on these tools, thus providing a promising way to obtain coverage  
\cite{molina2007functional}.}   

\blue{\ourtool{} detects both  \textit{b1} and \textit{b2} using the expression coverage of \cc{4} and \cc{6}, respectively. The expression coverage verifies that all the signals involved in the combinational logics \cc{4} and \cc{6} cover all possible values. One such combination will trigger the bugs  \textit{b1} and \textit{b2}, resulting in an incorrect output, which will be flagged as a mismatch. 
Thus, \ourtool{}’s coverage metrics aid detecting bugs \textit{b1} and \textit{b2}. }

In contrast to \ourtool{},
existing hardware fuzzers lose hardware intrinsic behaviors \blue{(e.g., floating wires, signal transitions)} while converting the target hardware into a software model~\cite{google_fuzzer},  
\blue{operate only on the select signals of the MUXes~\cite{laeufer2018rfuzz}},
operate only on the \blue{DFFs that determine the select signals of the MUXes}~\cite{difuzz}, or \blue{operate} at the protocol level~\cite{hyperfuzzing}. 
Hence, \blue{the coverage used by existing fuzzers will not be able to cover the bugs in \cc{4}, \cc{6}, and some DFFs in \cc{7} including the bugs we inserted, \textit{b1} and \textit{b2} (cf. Appendix~\ref{App:other_cov_metrics})}.

\subsection{Seed Generator}\label{sec:seed_gen}
Given that we have discussed the various coverage metrics to capture hardware behaviors, we now describe the seed generation in more detail. The seed generator generates \textit{seed inputs} that run on the DUT and are used to generate further inputs through mutation.

\noindent {\textbf{Seed inputs}.}
\ourtool{}'s goal is to detect software-exploitable vulnerabilities in the RTL model of the processors. 
Processors execute instructions using the data from the instruction memory.  
Hence, our fuzzer provides inputs at the Instruction Set Architecture (ISA) abstraction level by generating processor instructions.
The \textit{seed inputs} are data files containing a sequence of instructions, which are loaded onto the memory for execution. 

\noindent {\textbf{Instruction generator}}\label{sec:inst_generator}
generates the instructions for the \textit{seed inputs} from a set of valid instructions of the processor. 

\noindent {\textbf{Input format.}}
Each input consists of two types of instructions: \textit{configuration instructions} (CIs) and \textit{test instructions} (TIs). 
The CIs are needed to setup the baremetal environment, e.g., setting up the stack, exception handler table, and clearing the general-purpose registers. This baremetal environment allows \ourtool{} to run instructions directly on the processor without the need for an operating system.
The TIs are generated by the \textit{instruction generator}, which are the actual instructions used to fuzz the processor.
\subsection{Stimulus Generator}\label{sec:stimulus_generator}
The stimulus generator is responsible for mutating the current inputs, generating new inputs, and discarding the underperforming inputs.
Seed inputs are used to generate the first set of new inputs. 
We mutate the instructions directly as binary data instead of at a higher abstraction level such as assembly.
This allows us to mutate all the bits of the instruction based on the mutation technique used.
Thereby, we can test the processor with out-of-spec inputs like \textit{illegal} instructions (i.e., instructions not specified in the ISA) generated through mutation of the \textit{opcode bits} of the instruction.  
This allows us to detect issues that other verification techniques may not have detected, like the bug B3 in the \ariane{} and B8 in \orth{} processors, which cannot be detected with legal instructions. 

\noindent{\textbf{Mutation engine}}
performs the mutation operations on the instructions.
We mutate only the TIs since these are the instructions used to fuzz the processor. 
The CIs  are not mutated to ensure the correct initialization of the processor for fuzzing. 
The mutation techniques used by our fuzzer can be classified into two types.
The first type only mutates the data bits keeping the opcode unchanged.
These mutations increase the coverage on different data paths that are close to each other.
To generate bug-triggering out-of-spec inputs, the second type of mutation techniques mutates
both the data and the opcode bits.
Mutating the opcode bits will create inputs with new instruction sequences and help uncover different control paths in the DUT. 
This will help generate illegal instructions to test the processor with out-of-spec inputs.
We employ AFL-style mutation as detailed in Appendix~\ref{App:MutationTypes}.

Every time new inputs are generated by the stimulus generator, the code coverage data of these inputs is used to discard the under-performing inputs, thereby only retaining the inputs that trigger new code coverage points.
This helps steer the fuzzer towards discovering new coverage points quickly. 

\subsection{Optimization}\label{sec:our_fuzzing_tool}

\begin{figure}[tb]
    \centering
    \includegraphics[trim=40 20 20 20,clip,width=1\columnwidth]{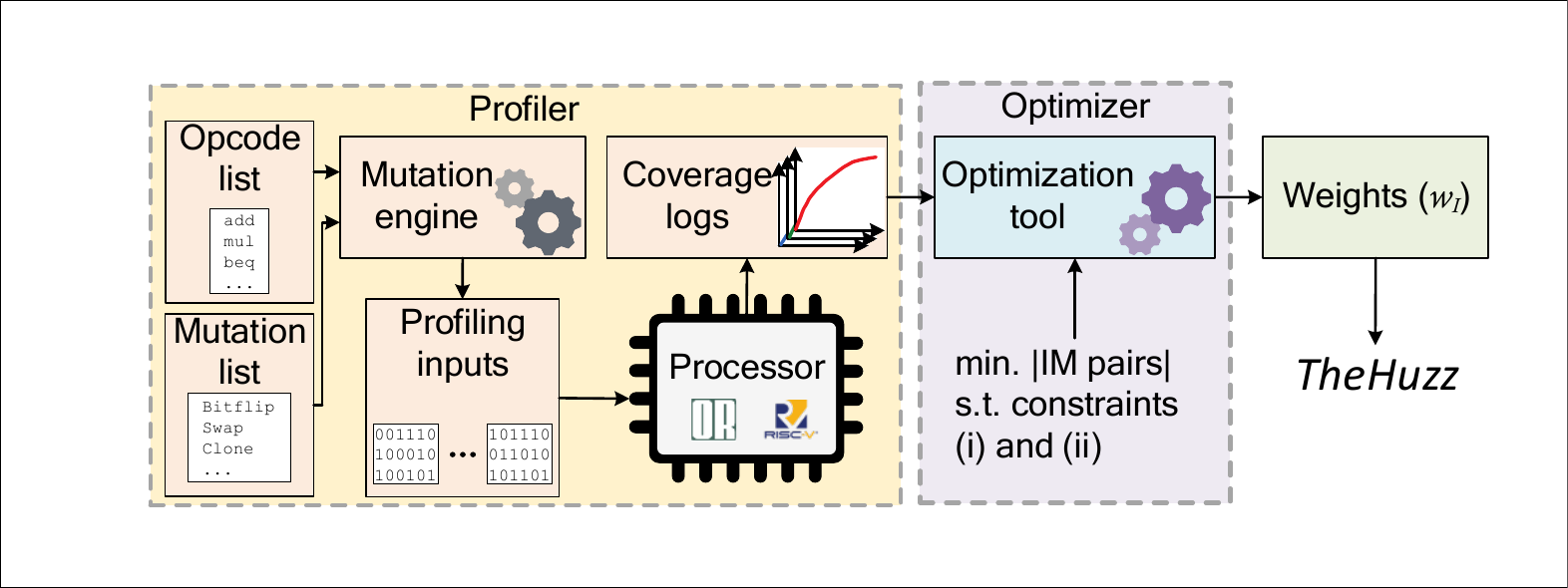}
    \caption{Optimization process used for \ourtool{}.}
    \label{fig:stress_testing}
    \vspace{-.35cm}
\end{figure}

We now propose an optimization for improving the efficiency of a processor fuzzer, as shown in Figure~\ref{fig:stress_testing}. 
Instead of using all the instructions and mutations, we optimally select the ones that achieve the best coverage. To this end, we first profile the individual instructions and mutations 
and formulate an optimization problem, which returns the optimal weights for each instruction-mutation (IM) pair. 

\noindent{\textbf{Profiler}} 
characterizes the control and data flow paths explored by each  IM pair. 
\ourtool{} generates the coverage values specific to each IM pair via hardware simulation. 

\noindent{\textbf{Optimizer}} 
aims to minimize the number of IM pairs while achieving the same amount of coverage as using all the IM pairs.
Let $I$ and $M$ be the sets of instructions and mutations, respectively.
Let $\mathcal{P}= I \times M$.  $\mathcal{C}$ denotes the union of coverage metrics such as statement, branch, expression, toggle, FSM, and condition. 
The coverage from the profiling phase for each IM pair is denoted by the indicator function $\mathcal{D}: \mathcal{P} \times \mathcal{C} \mapsto \{0, 1\}$. 
$\mathcal{C}_{d} \subseteq \mathcal{C}$ denotes the coverage points hit by an IM pair during the profiling phase. 
The optimization problem is to find the smallest subset of $\mathcal{P}$, denoted as $Q$, that covers all the coverage points identified during the profiling stage, $\mathcal{C}_{d}$. 
\red{Minimize $|Q|$ subject to constraints: (i)~$Q \subseteq P$ and (ii)~$\sum_{\{i,m\}\in Q} D((i,m),c) > 0, \hspace{0.2cm} \forall c \in C_{d}$.}
The optimizer returns the set $Q$ that contains the optimal IM pairs. \ourtool{} uses this information to generate the weights for each instruction-mutation pair $w_{(I,M)}(i,m) = \mathbb{I}_{\{i,m\}\in Q}, \hspace{0.2cm} \forall \ (i,m) \in P,$ where $\mathbb{I}$ is an indicator function.
The seed generator uses the weights, $w_I$,
to select instructions, and the stimulus generator uses the weights, $w_{M}$ to select the mutation techniques for each instruction and thereby, eliminating under-performing instructions and mutations.

\subsection{Bug Detection}
\label{subsec:bug_detection}
Software programs indicate bug triggers through crashes, memory leaks, and exit status codes. 
However, hardware intrinsically cannot provide such feedback because it does not crash or have memory leaks.
Thus, as performed in traditional hardware verification, we compare the outputs of GRMs and the DUT for the inputs generated by the fuzzer. 
Any \textit{mismatch} event indicates the presence of a bug, which is then manually analyzed to identify its cause. 
\section{Implementation}\label{sec:implementation}

We implemented \ourtool{} such that it is compatible with traditional IC design and verification flow, while effectively detecting security vulnerabilities.  
All the components are implemented in Python unless specified otherwise. 
We used CPLEX  for optimization~\cite{cplex}. 
\begin{table*}[!bth]
\centering
\caption{Bugs detected by \ourtool{}.}
\label{tab:bug_list_v1}
\resizebox{\textwidth}{!}{%
\begin{tabular}{|l|l|l|l|l|l|l|l|l|l|}
\hline
\multicolumn{1}{|c|}{\multirow{2}{*}{\textbf{Processor}}} &
  \multicolumn{1}{c|}{\multirow{2}{*}{\textbf{\begin{tabular}[c]{@{}c@{}}Prior work \\ using the \\ design\end{tabular}}}} &
  \multicolumn{2}{c|}{\textbf{Design size}} &
  \multicolumn{1}{c|}{\multirow{2}{*}{\textbf{Bug Description}}} &
  \multicolumn{1}{c|}{\multirow{2}{*}{\textbf{Location}}} &
  \multirow{2}{*}{\textbf{Coverage types}} &
  \multicolumn{1}{c|}{\multirow{2}{*}{\textbf{CWE}}} &
  \multicolumn{1}{c|}{\multirow{2}{*}{\textbf{\begin{tabular}[c]{@{}c@{}}New\\ bug?\end{tabular}}}} &
  \multirow{2}{*}{\textbf{\begin{tabular}[c]{@{}l@{}}\# instructions \\ to detect the bug\end{tabular}}} \\ \cline{3-4}
\multicolumn{1}{|c|}{} &
  \multicolumn{1}{c|}{} &
  \multicolumn{1}{c|}{\textbf{LOC}} &
  \multicolumn{1}{c|}{\textbf{\begin{tabular}[c]{@{}c@{}}Coverage \\ points\end{tabular}}} &
  \multicolumn{1}{c|}{} &
  \multicolumn{1}{c|}{} &
   &
  \multicolumn{1}{c|}{} &
  \multicolumn{1}{c|}{} &
   \\ \hline
\multirow{4}{*}{\begin{tabular}[c]{@{}l@{}}\ariane{} \cite{ariane_processor}\\ ISA: RISC-V \cite{riscv_home}\\ Design year: 2018\\ 64-bit, 6stage pipeline\end{tabular}} &
  \multirow{4}{*}{\begin{tabular}[c]{@{}l@{}}\cite{wistoff2020prevention, vsivsejkovic2020secure},\\ \cite{fischer2020hardware, dessouky2019hardfails},\\ \cite{openpiton_home}\end{tabular}} &
  \multirow{4}{*}{\blue{2.07 $\times 10^4$}} &
  \multirow{4}{*}{\blue{3.42 $\times 10^5$}} &
  \begin{tabular}[c]{@{}l@{}}\textbf{\ref{b1}}: Incorrect implementation of logic to detect \\ the FENCE.I instruction.\end{tabular} &
  Decoder &
  \blue{Branch} &
  CWE-440 &
  \greencmark{} &
  \blue{1.36 $\times 10^4$} \\ \cline{5-10} 
 &
   &
   &
   &
  \begin{tabular}[c]{@{}l@{}}\textbf{\ref{b2}}: Incorrect propagation of exception type in \\ instruction queue\end{tabular} &
  Frontend &
  \blue{Toggle} &
  CWE-1202 &
  \redxmark{} &
  \blue{4.02 $\times 10^4$} \\ \cline{5-10} 
 &
   &
   &
   &
  \textbf{\ref{b3}}: Some \textit{illegal} instructions can be executed &
  Decode &
  \blue{Condition} &
  CWE-1242 &
  \greencmark{} &
  \blue{1.81 $\times 10^6$} \\ \cline{5-10} 
 &
   &
   &
   &
  \textbf{\ref{b4}}: Failure to detect cache coherency violation &
  Cache controller &
  \blue{FSM} &
  CWE-1202 &
  \greencmark{} &
  \blue{1.72 $\times 10^5$} \\ \hline
\multirow{3}{*}{\begin{tabular}[c]{@{}l@{}}\morkx{} \cite{openrisc_home}\\ ISA: OpenRISC \cite{openrisc_home}\\ Design year: 2013\\ 32-bit, 6-stage pipeline\end{tabular}} &
  \multirow{3}{*}{\begin{tabular}[c]{@{}l@{}}\cite{deutschbein2018mining, zhang2018end},\\ \cite{krishnakumar2019msmpx, difuzz}\end{tabular}} &
  \multirow{3}{*}{\blue{2.21 $\times 10^4$}} &
  \multirow{3}{*}{\blue{4 $\times 10^4$}} &
  \begin{tabular}[c]{@{}l@{}}\textbf{\ref{b5}}: Incorrect implementation of the logic to \\ generate the \textit{carry} flag.\end{tabular} &
  ALU &
  \blue{Expression} &
  CWE-1201 &
  \greencmark{} &
  \blue{20} \\ \cline{5-10} 
 &
   &
   &
   &
  \begin{tabular}[c]{@{}l@{}}\textbf{\ref{b6}}: Read/write access checking not implemented \\ for privileged register\end{tabular} &
  Register file &
  \blue{Condition} &
  CWE-1262 &
  \greencmark{} &
  \blue{4.46 $\times 10^5$} \\ \cline{5-10} 
 &
   &
   &
   &
  \begin{tabular}[c]{@{}l@{}}\textbf{\ref{b7}}: Incomplete implementation of EEAR \\ register write logic\end{tabular} &
  Register file &
  \blue{Condition} &
  CWE-1199 &
  \greencmark{} &
  \blue{1.12 $\times 10^5$} \\ \hline
\multirow{3}{*}{\begin{tabular}[c]{@{}l@{}}\orth{} \cite{openrisc_home}\\ ISA: OpenRISC \cite{openrisc_home}\\ Design year: 2000\\ 32-bit, 5-stage pipeline\end{tabular}} &
  \multirow{3}{*}{\begin{tabular}[c]{@{}l@{}}\cite{zhang2017scifinder,gurumurthy2006automatic},\\ \cite{hicks2015specs,gurumurthy2007automatic},\\ \cite{khatri2019implementation,xuan2015configurable,bai201310gbps}\end{tabular}} &
  \multirow{3}{*}{\blue{3.16 $\times 10^4$}} &
  \multirow{3}{*}{\blue{3.90 $\times 10^4$}} &
  \textbf{\ref{b8}}: Incorrect forwarding logic for the GPR0 &
  Register forwarding &
  \begin{tabular}[c]{@{}l@{}}\blue{Condition} \\ \blue{and expression}\end{tabular} &
  CWE-1281 &
  \redxmark{} &
  \blue{174} \\ \cline{5-10} 
 &
   &
   &
   &
  \begin{tabular}[c]{@{}l@{}}\textbf{\ref{b9}}: Incomplete update logic of overflow bit for \\ msb \& mac instructions\end{tabular} &
  ALU &
  \blue{Toggle} &
  CWE-1201 &
  \greencmark{} &
  \blue{3.35 $\times 10^3$} \\ \cline{5-10} 
 &
   &
   &
   &
  \begin{tabular}[c]{@{}l@{}}\textbf{\ref{b10}}: Incorrect implementation of the logic to \\ generate the \textit{overflow} flag.\end{tabular} &
  ALU &
  \blue{Expression} &
  CWE-1201 &
  \greencmark{} &
  \blue{2.21 $\times 10^4$} \\ \hline
\begin{tabular}[c]{@{}l@{}}\blue{Rocket chip}~\cite{asanovic2016rocket}\\ \blue{ISA: RISC-V}~\cite{riscv_home}\\ \blue{Design year:2016} \\ \blue{32-bit, 5-stage pipeline}\end{tabular} &
  ~\cite{difuzz} &
  \blue{1.06 $\times 10^4$} &
  \blue{6.65 $\times 10^5$} &
  \begin{tabular}[c]{@{}l@{}}\textbf{\ref{b11}}: \blue{Instruction retired count not increased} \\ \blue{ when {\tt EBREAK}}\end{tabular} &
  \blue{Register file} &
  \blue{Condition} &
  \blue{CWE-1201} &
  \redxmark{} &
  \blue{776} \\ \hline
\end{tabular}%
}
\end{table*}

\noindent{\bf Register-Transfer-Level simulation.}
We simulate the target hardware using a leading industry tool, Mentor Graphics Modelsim~\cite{modelsim}.
This tool supports a wide variety of hardware description languages (HDLs) and  different hardware models: RTL, gate level, and transistor level.
We wrote custom Python scripts to process the logs of ModelSim to extract the coverage metrics---statement, branch, toggle, expression, and condition. 
It also generates instruction traces, which contain
the sequence of instructions executed along with the register or memory locations modified by each instruction and their updated values. 
Thus, \ourtool{} leverages existing hardware simulation tools to avoid instrumenting the HDLs. 

\noindent {\bf Seed generator} generates C programs that consist of configurations instructions (CIs) and test instructions (TIs). 
The CIs configure a baremetal C environment on the processors; we extract these CIs from the baremetal libraries of the corresponding ISAs, e.g., the RISC-V tests repository~\cite{riscv_tests}.
The TIs are the actual instructions used to fuzz the processor from the initial state.
\red{The TIs are generated uniformly from the ISA of the DUT.}
Each seed input has 20 TIs; this number is selected based on empirical observations before a random TI leads to a deadlock. 
Events like exceptions or instructions like branch, jump, system calls, and atomic instructions can cause the control flow of the processor to jump to a different location or even freeze for a large number of clock cycles, waiting for resources (in the case of atomic instructions).
\blue{The first half of the TIs are generated uniformly from the instructions that are less likely to trigger such events (e.g., arithmetic and logical instructions). This maximizes the number of TIs executed by the \ourtool{} in each simulation. The other half of the TIs are generated uniformly from all the instructions returned by the optimizer.}
Thus, the processor is reset after the execution of every 20 instructions and is simulated with new input. 
This results in periodical initialization of the processor control flow back to the location of the TI. 
The GCC toolchain compiles these C programs to generate the executable files 
which are loaded onto the processor RAM and used as seeds.

\noindent{\bf Stimulus generator}
 consists of the mutation and the feedback engines. 
The mutation engine mutates the TIs using the AFL-like mutations described in Appendix~\ref{App:MutationTypes}.
The feedback engine uses coverage logs for each mutated TI the from RTL simulation.
It retains the best performing instruction-mutation pairs and discards the ones that do not improve the coverage.

\noindent {\bf Golden Reference Models (GRMs).} 
\blue{We used   \textit{spike} ISA emulator~\cite{riscv_git_home} as GRMs for \ariane{}  and \rocketcore{},   and \textit{or1ksim} \cite{openrisc_github_home} as GRMs for \morkx{} and \orth{} processors. } 
\red{Ideally, GRMs that are designed specifically for the hardware are best suited to detect bugs. However, the processor designs we used for fuzzing are all open-source designs and do not have custom oracles. Hence, we used the ISA emulators, which are the closest to the processor models. }

\section{Evaluation}\label{sec:results}

We now describe the \red{three}\blue{four} open-source processors---\ariane{}, \morkx{}, \orth{}, and \blue{\rocketcore{}}---
used to evaluate our fuzzer \ourtool{} and present the evaluation results, along with bugs detected (see Table~\ref{tab:bug_list_v1}) and the coverage. 
\blue{We compare \ourtool{} with another fuzzer \difuzzrtl{}~\cite{difuzz}} and two traditional hardware verification techniques: random regression testing and formal verification\footnote{\blue{We did not compare with \rfuzz{} as it does not support processors~\cite{laeufer2018rfuzz}.}}.
The experiments are conducted on a 32-core Intel Xeon processor running at 2.6Ghz with 512GB of RAM with CentOS Linux release 7.3.1611. 

\subsection{Evaluation Setup}\label{sec:eval_setup}
With rich hardware-software interactions and complex hardware components, processor designs provide a challenging target for evaluating the potential of hardware fuzzers. While testing commercial processors is appealing, their closed-source nature makes Register-Transfer-Level (RTL) analysis impossible.
This is a challenge hardware researchers face, and hence, most papers which evaluate their tool's effectiveness on processors use open-source designs.
\blue{We have selected \red{three} four processors from two widely used open-source ISAs, OpenRISC~\cite{openrisc_home} and RISC-V~\cite{riscv_home}.} 
All these processors can run a modern Linux-based operating system.

\ariane{} (a.k.a. \textit{cva6} core)
is a RISC-V based, 64-bit, 6-stage, in-order processor, and supports a Unix-like operating system~\cite{ariane_processor}.
\red{Its developers and the open-source community have actively tested it for several years now. }
\morkx{} processor is a 32-bit OpenRISC based processor. From the three possible configurations, we selected the 6-stage \textit{Cappuccino} configuration, as it is the most complex design. 
Developers and the open-source community have evaluated this design for more than seven years. 
 \orth{} processor is a 32-bit OpenRISC based processor and is one of the first open-source processors used for more than two decades~\cite{openrisc_home}.
\blue{\rocketcore{} is a RISC-V based, 64-bit, 5-stage, in-order scalar processor, and supports a Unix-like operating system~\cite{asanovic2016rocket}. 
}
\red{OpenRISC instruction set architecture (ISA), one of the first popular open-source ISAs, is used in commercial applications like Samsung DTV System on Chips (SoCs), A31 ARM-based SoCs, and NASA's TechEdSat satellite~\cite{openrisc_wiki}. 
Nowadays, RISC-V has become one of the most popular open-source ISA for academic and industrial processors~\cite{riscv_home}.}
RISC-V open-source processors are widely used in prior work in hardware verification and security, as shown in Table~\ref{tab:bug_list_v1}, and have proven to be effective replacements for commercial designs.

\vspace{-0.3cm}
\subsection{Bugs Detected} \label{sec:bugs}
We now detail the vulnerabilities detected by \ourtool{}. 
We found 8 new bugs. We map each bug to the relevant hardware Common Weakness Enumerations (CWEs), as listed in Table~\ref{tab:bug_list_v1}.
We present bugs \ref{b1}, \ref{b4}, and  \ref{b6} in detail as we exploit them in Section~\ref{sec:exploits}, Appendix~\ref{App:bugsFound} details the other bugs.

\begin{lstlisting}[language=Verilog, label = {listing:ariane_bug_fence}, caption={Verilog code snippet for \ref{b1} in \ariane{}.},style=prettyverilog,float,belowskip=-15pt,aboveskip=-5pt]
// Memory ordering instructions
riscv::OpcodeMiscMem: begin
  instruction_o.fu  = CSR;
  instruction_o.rs1 = '0;
  instruction_o.rs2 = '0;
  instruction_o.rd  = '0;
  case (instr.stype.funct3)
    // FENCE: Currently implemented as a whole DCache flush boldly ignoring other things
    3'b000: instruction_o.op  = ariane_pkg::FENCE;
    // FENCE.I
    3'b001: begin
      if (instr.instr[31:20] != '0)
        illegal_instr = 1'b1;
      instruction_o.op  = ariane_pkg::FENCE_I;
    end
    default: illegal_instr = 1'b1;
  endcase
  if (instr.stype.rs1 != '0 || instr.stype.imm0 != '0 || instr.instr[31:28] != '0)
    illegal_instr = 1'b1;
end
\end{lstlisting}
\subsubsection{\textbf{Bugs in \ariane{} Processor}}\label{sec:ariane_bugs}
\noindent\begin{bug_ns}[\label{b1}] 
is located in the decode stage of \ariane{}.
According to the RISC-V specification~\cite{riscv_home}, the decoder should ignore 
certain fields in a \textit{FENCE.I} instruction, which  
enforces cache coherence in the processor (e.g., by flushing the instruction cache and instruction pipeline).  It also ensures 
that the correct instruction memory is used for execution when performing memory sensitive operations (e.g., updating the instruction memory).
The bug is that the decoder \red{in \ariane{} implementation} does not ignore the \textit{imm} and \textit{rs1} fields and expects a value of \textit{0} in these fields, as seen in Lines 12 and 18 of Listing~\ref{listing:ariane_bug_fence}. 
This \ariane{} implementation declares valid instructions as illegal (Lines 13 and 19) \blue{due to} this additional constraint on the \textit{imm} and \textit{rs1} fields, thus violating the specification. 
We detected this bug when the fuzzer generated a \textit{FENCE.I} instruction with a non-zero value in the \textit{imm} field. 
\ariane{} raised an exception saying the instruction is \textit{illegal}, whereas the oracle \textit{spike} successfully executed the instruction, resulting in a mismatch\footnote{  
We refer to these \textit{FENCE.I} instructions that \ariane{} fails to detect as \textit{failing-FENCE.I} and the rest as the \textit{working-FENCE.I}.}.
Due to this bug, \textit{failing-FENCE.I} will not be executed, resulting in a potential violation of cache coherence. 
This bug is similar to the expected behavior violation vulnerability, CWE-440~\cite{hardware_cwe}.
\end{bug_ns}

\begin{bug}[\label{b2}] 
is in the instruction queue of the frontend stage of \ariane{}. 
The bug is that a fixed exception is forwarded without the actual exception. We detected this bug as a mismatch in the value of a register that loads the exception type when an exception occurs. 
Operating systems that assume that instruction access-faults are raised correctly will not behave as expected, and triggering this bug may lead to undefined (and possibly exploitable) behavior.
Also, an incorrect exception handling might be executed, resulting in a memory and storage vulnerability, CWE-1202~\cite{hardware_cwe}. 
\end{bug}

\begin{bug}[\label{b3}] 
is that the decode stage does not correctly check for certain illegal instructions.
It was detected as a mismatch when the fuzzer generated one such illegal instruction. 
Due to this, any undocumented instruction of  a certain value 
can be executed on \ariane{}, resulting in an undocumented feature vulnerability, CWE-1242~\cite{hardware_cwe}.
\end{bug}

\begin{bug}[\label{b4}]  
According to the RISC-V specification~\cite{riscv_home}, when the instruction memory is modified, the software should handle cache coherency using \textit{FENCE.I} instruction.
Failure to handle cache coherency results in undefined behavior, wherein processors may use stale data and\red{ leading to an} incorrect execution of instructions~\cite{sorin2011primer}.
When the fuzzer generated an input program that modified the instruction memory but did not use a \textit{FENCE.I} instruction, \ourtool{} detected a mismatch in the trace logs of \ariane{} and \textit{spike}. 
This mismatch could have been avoided if the RISC-V specification or the \ariane{} processor\red{ checked to} detect\blue{ed} violations of cache coherency in hardware.
Because of this bug, software running on \ariane{} could run into cache coherency issues and remain undetected if the \textit{FENCE.I} instruction is used incorrectly, resulting in a memory and storage vulnerability, CWE-1202~\cite{hardware_cwe}.
The \ariane{} exploit in Section~\ref{sec:ariane_exploit} this bug and bug B1 to successfully exploit a theoretically safe program. 

\end{bug}
\subsubsection{\textbf{Bugs in \morkx{}{} Processor}}
\noindent\begin{bug_ns}[\label{b5}] 
is the inaccurate implementation of the \textit{carry} flag logic for subtract operations.
The fuzzer generated inputs that triggered this bug by mutating the data bits of subtract instructions. 
This caused a mismatch in the value of the \textit{carry} flag 
between the RTL and golden reference model (GRM). 
This bug can cause incorrect computations, including those used in cryptographic functions, resulting in corruption and compromise of the processor security (CWE-1201~\cite{hardware_cwe}).

\end{bug_ns}

\begin{bug}[\label{b6}]  
The register file stores, updates, and shares the value of all the architectural registers.
These registers include the general- and special-purpose registers (GPRs and SPRs, respectively). 
\textit{Read} and \textit{write} operations to the SPRs are restricted based on the privilege mode of the processor,  as per the OpenRISC specification~\cite{openrisc_home}. 
The Exception Program Counter Register (\texttt{EPCR}) is an SPR that stores the address to which the processor should return after handling an exception. A user-level program should not be able to access this register.
The bug in \morkx{} is that the register file does not check for privilege mode access permissions when performing \textit{read} and \textit{write} operations on \texttt{EPCR}. 
This bug was detected when our fuzzer generated an instruction that tried to write into \texttt{EPCR} from user privilege mode. 
Due to this bug, an attacker can write into \texttt{EPCR} from user privilege mode and control the return address of the processor after handling an exception (CWE-1262~\cite{hardware_cwe}). 
This bug can have severe security consequences like privilege escalation, as demonstrated in our \morkx{} exploit in Section~\ref{sec:mor1kx_exploit}. 
\end{bug}
\begin{bug}[\label{b7}]. 
The register file in \morkx{} does not allow one to write into the Exception Effective Address Register (\texttt{EEAR}), even for supervisor privilege mode.
This bug is detected when our fuzzer generated an instruction that tried to write into EEAR from the supervisor privilege mode. 
This bug prevents programs from updating \texttt{EEAR}, resulting in incorrect executions.  
Thus, it prevents software from correctly performing exception handling.
This bug is similar to CWE-1199~\cite{hardware_cwe}. 
\end{bug}

\subsubsection{\textbf{Bugs in \orth{} Processor}}
\noindent\begin{bug_ns}[\label{b8}] 
 is that the register forwarding logic forwards a non-zero value for \texttt{GPR0} if a previous instruction in the pipeline writes to \texttt{GPR0}. 
We found this bug as a mismatch when the fuzzer applied an ADD instruction to create a data hazard for {\tt GPR0}.
This bug can result in incorrect computations since \texttt{GPR0} is frequently used by software to check for conditions. 
An attacker can cause data hazards to obfuscate the behavior of malware, e.g., by jumping to an offset computed by an instruction that uses \texttt{GPR0}.
This bug is similar to CWE-1281~\cite{hardware_cwe}, where a sequence of processor instructions resulting in unexpected behavior.
\end{bug_ns}

\begin{bug}[\label{b9}]  
 is that the \textit{overflow} flag is not correctly calculated for multiply and subtract (MSB) or the multiply and accumulate (MAC) instructions. 
This bug results in the failure of the software programs to detect the \textit{overflow} events. 
Thus this bug is a core and compute issue vulnerability, CWE-1201~\cite{hardware_cwe}, resulting in more software vulnerabilities.
\end{bug}

\begin{bug}[\label{b10}]  
 is the incorrect overflow logic for the subtract instruction. 
The bug was detected 
when the fuzzer was mutating data bits of subtract instruction. 
This bug also compromises the security mechanisms relying on the \textit{overflow} flag and is a core and compute issue vulnerability, CWE-1201~\cite{hardware_cwe}. 

\end{bug}

\subsubsection{\textbf{Bugs in \rocketcore{} Processor}}
\begin{bug}[\label{b11}] \blue{ is that the instruction retired count does not increase on an {\tt EBREAK} instruction. It was detected when the fuzzer executed the {\tt EBREAK} instruction.} 
\blue{\ourtool{} was able to detect the only bug, \ref{b11} reported by \difuzzrtl{}. 
\ourtool{} detects using only 776 instructions and is 6.7$\times$ faster than \difuzzrtl{}}.  
\end{bug}

All the bugs except for \ref{b2}\blue{,}\red{and} \ref{b8}\blue{, and \ref{b11}} are new bugs detected by \ourtool{}. 
\ref{b2} is fixed in the latest version of Ariane. 
\ref{b8} is first reported in~\cite{zhang2018end}.

\subsection{Case Study: Exploitability}\label{sec:exploits}
We now present the two exploits we crafted to demonstrate the security implications of the bugs found by \ourtool{}. 
\red{\ref{b1} and \ref{b4} can be leveraged in a return-oriented programming~(ROP) exploit in the \ariane{} processor, and \ref{b6} leads to a privilege escalation exploit in the \morkx{} processor.}
Both attacks can be mounted from unprivileged software.

\subsubsection{\textbf{Ariane FENCE.I Exploit}}\label{sec:ariane_exploit}

\blue{
The Ariane exploit leverages \ref{b1} and \ref{b4} to cause incoherence in the instruction cache.
As a result, in the contrived ``safe'' just-in-time (JIT) compiler we developed to demonstrate this bug, an attacker can generate inputs that selectively invalidate cache lines containing old instructions.
This program uses an extension of the \textit{FENCE.I} instruction which should fall back to standard fence behavior and flush the entire instruction cache as the extension is not understood by Spike or Ariane.
An attacker first loads a region of executable code (which does not contain a vulnerability) into the cache by executing it.
The attacker then overwrites the same region of executable code with new instructions (which also does not contain a vulnerability), then executes separate code which jumps to instructions which align to cache lines the attacker wishes to invalidate.
After, they execute the original region of executable code, at which point the behavior of Spike and Ariane diverge.
In Spike, the new instructions will be present and will execute as expected with no vulnerabilities present.
This is because Spike successfully identified the \textit{FENCE.I} instruction, but did not recognise its extension, and fell back to flushing the entire cache.
In Ariane, the old instructions will be present; Ariane fails to recognise the \textit{FENCE.I} instruction as it instead marks it as an illegal instruction, an implementation which is non-compliant with the RISCV ISA.
Because the cache lines were only invalidated in regions selected by the attacker, the attacker is able to successfully replace bounds checks in the original program with effectively nops, leading to a vulnerability which was neither present in the old or the new JIT code.
As a result, the attacker is able to inject a stack overflow vulnerability and gain arbitrary code execution.
A more detailed description of the vulnerability, exploit, ramifications, and threat model are presented in Appendix~\ref{App:mor1kx_exploit}.
}

\begin{table*}[t!]
\centering
\caption{Hardware complexity encountered while using industry-standard JasperGold~\cite{jasperGold} to detect the bugs.}
\label{tab:bugStatsV3}
\resizebox{\textwidth}{!}{%
\begin{tabular}{|c|l|l|l|l|l|l|l|l|l|l|l|}
\hline
\textbf{Processor}                 & \multicolumn{4}{c|}{\textbf{\ariane{}}}   & \multicolumn{3}{c|}{\textbf{\morkx{}}} & \multicolumn{3}{c|}{\textbf{\orth{}}} & \rocketcore{}\\ \hline
\diagbox{\textbf{Statistics}}{\textbf{Bug}} & \ref{b1} & \ref{b2} & \ref{b3} & \ref{b4} & \ref{b5}    & \ref{b6}    & \ref{b7}   & \ref{b8}   & \ref{b9}   & \ref{b10}  & \ref{b11} \\ \hline
\textbf{No. of modules} & 1   & 9   & 1   & 655 & 1   & 4   & 4   & 6   & 1   & 1 & 1  \\ \hline
\textbf{No. of inputs}  & 518 & 627 & 518 & 3   & 298 & 752 & 752 & 703 & 123 & 123 & 284 \\ \hline
\textbf{No. of states}             & 2.51e+58 & 2.16e+68 & 2.16e+68 & 2.01e+59 & 4.72e+10    & 1.55e+11    & 1.55e+11   & 3.83e+11   & 1.29e+10   & 1.29e+10  & 2.23e+20 \\ \hline
\end{tabular}%
}
\end{table*}

\subsubsection{\textbf{mor1kx EPCR Register Exploit}} \label{sec:mor1kx_exploit}
The mor1kx exploit \blue{leverage}s the \ref{b6} to set the exception program counter register (\texttt{EPCR}) to point to an attacker-controlled exploit function.
An exception return instruction is executed to mimic the return from an exception event, causing the processor to update the program counter (\texttt{PC}) and status register (\texttt{SR}) values with \texttt{EPCR} and exception status register (\texttt{ESR}) values, respectively. \red{, as defined in the specification.}
The SR stores the privilege level.
By performing the exploit when the \texttt{ESR} stores a higher-privilege level, execution jumps to the exploit function while overwriting the privilege level stored in \texttt{SR}. 
Thus, we successfully achieved privilege escalation in the \morkx{} processor.    
Appendix \ref{App:mor1kx_exploit} explains this exploit in detail.

\subsection{Coverage Analysis}
\label{sec:coverageAnalysis}

\begin{figure}[tb!]
    \centering
    \hspace*{0cm}
    \includegraphics[trim=20 20 20 20,clip,width=1\columnwidth]{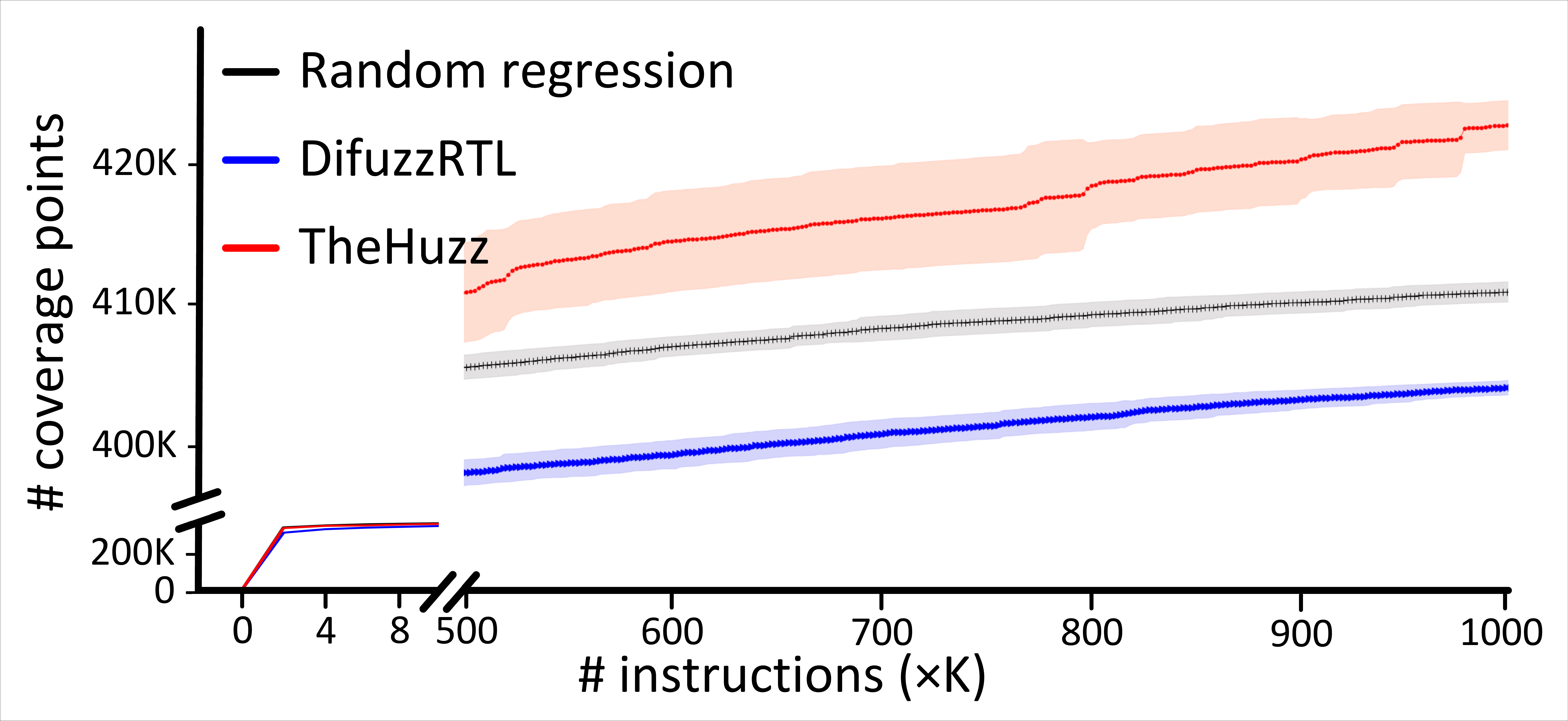}
    \caption{\blue{Coverage analysis of random regression testing, \difuzzrtl{}, and \ourtool{} for the Rocket core processor. }
    }
    \label{fig:total_cov_difuzz}
    \vspace{-.35cm}
\end{figure}

Figure~\ref{fig:total_cov_difuzz} shows the coverage achieved by random regression testing, \difuzzrtl{}, and \ourtool{} for the Rocket core processor. \blue{Each experiment is repeated 10 times.}
Even after \red{100 hours}\blue{1M instructions}, \blue{both} random regression testing \blue{and \difuzzrtl{} } \red{does}\blue{did} not improve \red{its}\blue{their} coverage beyond \blue{2.5}\% than what \red{it has}\blue{they }  \red{achieved}\blue{collected} after applying \red{17 hours}\blue{300K instructions}; on the other hand \ourtool{}'s coverage kept increasing. 
\ourtool{} is slower in the beginning \blue{than random regression testing} as the fuzzer uses a set of instructions until it cannot reach new coverage points; in that case, it discards and selects new a set of instructions. 
\blue{\ourtool{} achieved the 404.1K coverage points achieved by \difuzzrtl{} at 3.33$\times$ the speed of \difuzzrtl{}. 
\ourtool{} and random regression testing outperformed \difuzzrtl{} because \difuzzrtl{} is guided by the control-register coverage, which does not capture many hardware behaviors (cf. Appendix~\ref{App:other_cov_metrics}).} 
The p-value from the Mann-Whitney U test~\cite{navidi2008statistics} shows that the result is statistically significant (p < 0.05) with a p-value of 
\blue{1.4e-4 for both random regression testing and \difuzzrtl{}}.
\blue{The Vargha-Delaney A12 measure returned \ourtool{} as the best performing technique when compared with random regression testing and \difuzzrtl{}, 
}

\blue{The instrumentation overhead of \difuzzrtl{} is 18\% in terms of lines of Verilog code. \ourtool{} does not instrument Verilog code explicitly and instead relies on the commercial tools which do not produce the overhead information. Hence, the instrumentation overheads of these two fuzzers are not comparable.} 
\blue{The runtime overhead for \ourtool{} (71\%) is greater than \difuzzrtl{} (6.9\%) since \ourtool{} requires accessing multiple files to collect all the coverage, whereas \difuzzrtl{} only needs to collect control-register coverage.}

\subsection{Comparison with Formal Verification}
\label{sec:formalComparison}
\red{Apart from comparing with random regression testing,} 
We also compared our fuzzer with another standard approach used by the semiconductor industry---formal verification. 
For this purpose, we used the industry-leading formal verification tool, Cadence JasperGold~\cite{cadence_home}.
However, there are two challenges in performing this comparison. 
First, there is no industry-standard formal tool that can produce a set of instructions that can trigger a hardware bug in RTL, even if the bug is known apriori. 
Second, these industry tools require one to write assertions targeting each  vulnerability manually. 
Thus, the usage of formal tools in this scenario requires one to know of these vulnerabilities apriori---unlike \ourtool{}, which does not make any such assumptions. 

To manually write these assertions, one has to know the entire design, identify the signals and specific conditions that trigger the security vulnerability. 
This step is highly cumbersome given the vast number of modules, signals, and states in processors, as shown in Table~\ref{tab:bugStatsV3}.
Many bugs are cross-modular, and hence, they require one to load multiple modules, which only makes writing assertions difficult as they now need to consider signals across modules and their interactions. 
These tools only produce Boolean assignments to the inputs of these modules and not a set of instructions that violate these assertions. 
As shown Table~\ref{tab:bugStatsV3}, the number of inputs ranges in few hundreds, thereby increasing the number of states that need to be checked, leading to state-space explosion. 
Some bugs like \ref{b4} require one to load the entire system-on-chip into the formal tool, which is not always feasible due to state-space explosion. 
\red{Additionally, identifying a set of instructions that assign the desired values to those many signals is not a straightforward task and currently, no tool exists for this purpose.}
Thus, in contrast to \ourtool{}, existing formal tools are
resource intense, error-prone, and not scalable to complex bugs and larger designs, apart from  relying heavily on human expertise and prior knowledge of hardware vulnerabilities. 
\section{Related Work}\label{sec:related_work}
\begin{table*}[bth!]
\centering
\caption{\blue{Comparison with the prior work on hardware fuzzers.}}
\label{tab:related_work_comparison}
\resizebox{\textwidth}{!}{%
\begin{tabular}{|l|l|l|l|l|l|l|l|l|l|l|l|}
\hline
\multicolumn{1}{|c|}{\textbf{Methodology}} &
  \multicolumn{1}{c|}{\textbf{Fuzzer used}} &
  \multicolumn{1}{c|}{\textbf{HDL}} &
  \multicolumn{1}{c|}{\textbf{Simulator}} &
  \multicolumn{1}{c|}{\textbf{\begin{tabular}[c]{@{}c@{}}\blue{Target }\\\blue{design}\end{tabular}}} &
  \multicolumn{1}{c|}{\textbf{\begin{tabular}[c]{@{}c@{}}Design \\ knowledge\end{tabular}}} &
  \multicolumn{1}{c|}{\textbf{\begin{tabular}[c]{@{}c@{}}Largest \\ design \\ (Lines of code)\end{tabular}}} &
  \multicolumn{1}{c|}{\textbf{Metrics used}} &
  \multicolumn{1}{c|}{\textbf{\begin{tabular}[c]{@{}c@{}}Comparison \\ against random \\ regression testing\end{tabular}}} &
  \multicolumn{1}{c|}{\textbf{\begin{tabular}[c]{@{}c@{}}Bugs \\ reported\end{tabular}}} &
  \multicolumn{1}{c|}{\textbf{\begin{tabular}[c]{@{}c@{}}Exploitable \\ from \\ software\end{tabular}}} &
  \multicolumn{1}{c|}{\textbf{\begin{tabular}[c]{@{}c@{}}Exploits \\ presented\end{tabular}}} \\ \hline
\begin{tabular}[l]{@{}l@{}}RFUZZ\\ \cite{laeufer2018rfuzz}\end{tabular} &
  H/W fuzzer &
  FIR-RTL &
  Any &
  \begin{tabular}[c]{@{}l@{}}\blue{RTL }\\ \blue{designs}\end{tabular} &
  Not required &
  \begin{tabular}[c]{@{}l@{}}5-stage Sodor \\ core (4,088)\end{tabular} &
  \begin{tabular}[c]{@{}l@{}}\blue{mux-coverage} \\ \red{coverage}\end{tabular} &
  \begin{tabular}[c]{@{}l@{}}$\sim$5\% increase in \\mux coverage\end{tabular} &
  0 &
  N/A &
  0 \\ \hline
\begin{tabular}[l]{@{}l@{}}Hyperfuzzing\\\cite{hyperfuzzing}\end{tabular} &
  \begin{tabular}[c]{@{}l@{}}S/W AFL \\ fuzzer\end{tabular} &
  Any &
  Verilator &
  \begin{tabular}[c]{@{}l@{}}\blue{SoC }\\ \blue{designs}\end{tabular} &
  \begin{tabular}[c]{@{}l@{}}Need \\ security rules\end{tabular} &
  \begin{tabular}[c]{@{}l@{}}SHA crypto \\ engine (1,196)\end{tabular} &
  None &
  N/A &
  0 &
  N/A &
  0 \\ \hline
\begin{tabular}[l]{@{}l@{}}Trippel et al.\\\cite{google_fuzzer}\end{tabular} &
  \begin{tabular}[c]{@{}l@{}}S/W AFL \\ fuzzer\end{tabular} &
  Any &
  Verilator &
  \begin{tabular}[c]{@{}l@{}}\blue{RTL }\\ \blue{designs}\end{tabular} &
  Not required &
  KMAC (4,585) &
  \begin{tabular}[c]{@{}l@{}}FSM\red{ coverage} \blue{, line, edge, toggle,} \\ \blue{and functional coverage}\end{tabular} &
  \begin{tabular}[c]{@{}l@{}}Two orders magnitude\\ faster for datapath FSMs\end{tabular} &
  0 &
  N/A &
  0 \\ \hline
\begin{tabular}[l]{@{}l@{}}DIFUZZRTL\\\cite{difuzz}\end{tabular} &
  H/W fuzzer &
  Any &
  Any &
  \begin{tabular}[c]{@{}l@{}}\blue{CPU }\\ \blue{designs}\end{tabular} &
  Not required &
   \begin{tabular}[c]{@{}l@{}}Boom\\(12,956 in Scala)\end{tabular} &
  \begin{tabular}[c]{@{}l@{}}Control-register \\ coverage\end{tabular} &
  \begin{tabular}[c]{@{}l@{}}$\sim$10\% increase in \\control-register coverage\end{tabular} &
  16 &
  Not reported &
  0 \\ \thickhline  
\textit{\textbf{\textit{\bf \ourtool{}}}} &
  H/W fuzzer &
  Any &
   \begin{tabular}[c]{@{}l@{}}\blue{Commercial},  \\ \blue{industry-standard}\\ \blue{HDL simulator}\end{tabular} &
  \begin{tabular}[c]{@{}l@{}}\blue{CPU }\\ \blue{designs}\end{tabular} &
  Not required &
  Ariane (20,698) &
  \begin{tabular}[c]{@{}l@{}}\red{Line}\blue{statement}, toggle\red{ (FSM)},  \\ branch, expression, \red{and}\\ \red{statement}\blue{condition, and FSM} coverage\end{tabular} &
  \begin{tabular}[c]{@{}l@{}}\blue{2.86}\% increase \\ in code coverage \\ metrics\end{tabular} &
  10 &
  Yes &
  2* \\ \hline
\end{tabular}%
}
\begin{footnotesize}\blue{*In theory, the bugs discovered can be used to build more than two exploits, but we show only two due to page limitations.} \end{footnotesize}
\end{table*} 

We now describe the limitations of the existing attempts to fuzz hardware and how \ourtool{} is different from them, as summarized in Table~\ref{tab:related_work_comparison}. 

\noindent\textbf{\rfuzz{}} is a mux-coverage-guided fuzzer for hardware designs~\cite{laeufer2018rfuzz}. 
\blue{Although this technique can fuzz designs on FPGAs, 
it is computationally intensive and does not scale to large designs~\cite{difuzz}.
Additionally, its coverage metric does not capture many hardware behaviors (see Appendix~\ref{App:other_cov_metrics}). 
It is also ineffective in finding any bugs. 
}

\noindent\textbf{HyperFuzzing} proposes a new grammar to represent the security specification rules for hardware, converts the hardware design into equivalent software models, and fuzzes them using AFL fuzzer~\cite{hyperfuzzing}.  
It is inapplicable to general hardware designs like finite state machines (FSMs) or combinational logic and 
requires a lot of human intervention, including writing security specifications manually. 
It did not report any bugs. 

\noindent\textbf{Fuzzing hardware like software} 
translates the hardware design to software models and fuzzes them using a software fuzzer~\cite{google_fuzzer}. 
\blue{While this is a promising approach,
it is limited by the strength of existing open-source tools (i.e., Verilator~\cite{verilator}): they currently do not support many constructs of HDLs such as latches, floating wires, etc. 
It did not report any bugs.
}
The largest benchmark used by this technique has 4,585 lines of code (LOC).  It also does not scale to real-world designs like processors.
For instance, while fuzzing Google's OpenTitan SoC~\cite{opentitan}, this work could only fuzz the peripheral modules but not the {\it iBex} processor\footnote{We did not fuzz the iBex processor as it does not have an independent GRM and does not support other software emulators as GRMs.} 
in it. 

\noindent\textbf{\difuzzrtl{}}, a recent work, uses a custom-developed control-register coverage as feedback for the fuzzer by instrumenting the HDL~\cite{difuzz}. 
The technique only focuses on the FSM coverage and does not check for \red{statement, }toggle, expression\red{, and condition}\blue{, and FSM} coverage points, thereby missing the bugs in \cc{3}, \cc{4}, and floating wires in \red{\cc{3}}\blue{\cc{5}} in  Figure~\ref{fig:ExampleDesign} \blue{(see Appendix~\ref{App:other_cov_metrics} for more details)}. 
None of the bugs found by this fuzzer are shown to be exploitable, as most bugs are triggered by physically controlling the interrupt signals with precise timing; such interrupt signals are not usually exposed to unprivileged software~\cite{ariane}. 
The fuzzer is also slower in detecting the bugs as it compares the processor state after the entire program is executed, while our fuzzer performs comparison after each instruction is executed. 

In contrast, {\bf \ourtool{}}: 
(i)~is compatible with traditional IC design verification flow allowing for seamless integration by using coverage metrics already widely used in the semiconductor industry,
(ii)~is scalable to large, complicated, industrial-designs with several tens of thousands of code, and not just small FSM designs;
(iii)~captures many intrinsic hardware behaviors, such as signal transitions and floating wires, using multiple coverage metrics: statement, toggle, branch, expression, condition, and FSM;
(iv)~does not require the designer to specify security rules, and
(v)~detects several bugs that lead to severe security exploits. 
Instead, we compare how the software views the hardware (i.e., ISA emulator) and how the hardware actually behaves (i.e., Verilog), leading to an effective hardware fuzzer. 

\blue{\section{Discussion and Limitations}\label{sec:discussion}}

\noindent\textbf{Requirement of Golden Reference Models (GRMs).}
\blue{
\ourtool{}'s and other hardware fuzzers'~\cite{difuzz, hyperfuzzing} depend on GRMs to find vulnerabilities.  
Such GRMs are widely available in the semiconductor industry. 
Verification of many commercial (proprietary and open-source) CPUs critically depend on the availability of GRMs, including many industrial, large-scale designs, e.g. Intel x86 Archsim \cite{intel_archsim}, AMD x86 Simnow \cite{amd_simnow}, ARM Cortex Neoverse \cite{arm_cortex_neoverse}, and ARM Fast Models \cite{arm_fast_models}.
        Thus, the reliance on GRM is not a limiting factor for \ourtool{}.}
       \blue{
       Sometimes, the GRM itself can be buggy, thereby causing false positives. This situation is highly unlikely because GRMs are carefully curated and versioned with legacy code, and rigorously tested.
       Verifying a GRM  is easier as it is written at a higher abstraction level and is thus less complex than an RTL model.}

\noindent\textbf{Requirement of Register Transfer-Level (RTL) source code.}
 \blue{\ourtool{} depends on RTL access, similar to previous works such as \difuzzrtl{}~\cite{difuzz}, \rfuzz{}~\cite{laeufer2018rfuzz}, and Hyperfuzzing~\cite{hyperfuzzing}. As mentioned in Section~\ref{sec:HSDL}, verification teams already have access to RTL. 
 An attacker can also buy RTL models of the target design, as many companies like Imagination Tech. Limited \cite{imagination}, Cadence \cite{cadence_ip}, Synopsys   \cite{synopsys_ip} sell proprietary hardware designs, and run \ourtool{} on them as  these designs are compatible with industry-standard tools. While companies like Intel and ARM do not reveal the RTL model of their processors, attackers can use reverse engineering services from companies like TechInsights \cite{TechInsights} on  the target chip and use gate-level to RTL reverse engineering techniques~\cite{subramanyan2013reverse} to obtain the RTL model. }
 
\noindent\textbf{FPGA emulations.}
\blue{\difuzzrtl{} and \rfuzz{} can fuzz processors faster through FPGA emulation than RTL simulations~\cite{difuzz, laeufer2018rfuzz}. 
\ourtool{} uses the coverage metrics implemented by EDA simulation tools like Modelsim~\cite{modelsim} and Synopsys VCS~\cite{synopsys_vcs}. 
These coverage metrics are not readily available for FPGA emulations, thereby limiting \ourtool{}’s applicability to fuzz FPGA-emulated designs. }

\noindent\textbf{Fuzzing non-processor designs.}
\blue{Currently, \ourtool{}, similar to \difuzzrtl{}~\cite{difuzz}, 
is limited to fuzzing processor designs since it generates processor specific inputs.
        These fuzzers cannot fuzz standalone hardware components like SoC peripherals, memory modules, and other hardware accelerators, which are targeted by \rfuzz{} and Tripple et al.~\cite{google_fuzzer}.
        \ourtool{} could be extended to fuzz non-processor designs by fuzzing the individual input signals of the design.
        The seeds would be assignments to individual input signal values rather than instructions.
        The coverage metrics and the bug detection mechanism used by \ourtool{} will still be applicable.} 

\noindent\textbf{Fuzzing parametric properties of hardware.}
\blue{\ourtool{} currently fuzzes only processors for functional behavior but not for parametric behavior (e.g., cache timing behavior) and thereby cannot detect side-channel vulnerabilities.  One can extend \ourtool{} to cover such vulnerabilities by developing timing-related coverage properties and targeting them.  
}

\vspace{0.3in}
\section{Conclusion}

Bugs in hardware are increasingly exposed and exploited. Current techniques fall short of detecting bugs, as our results demonstrated by finding bugs in a 20-year old processor and others. This calls for a revamp of security evaluation methodologies for hardware designs. 

We presented an instruction fuzzer, \ourtool{}, for processor-based hardware designs. 
The effectiveness of \ourtool{} is shown by fuzzing three popular open-sourced processor designs. 
\ourtool{} has detected eight new bugs in the three designs tested and three previously detected bugs.
These bugs, when used individually or \red{stitched together}\blue{in tandem}, resulted in ROP and privilege escalation exploits that could compromise both hardware and software, as shown in the two exploits we presented. 
\blue{Our fuzzer achieved 1.98$\times$ and 3.33$\times$ the speed compared to the industry-standard random regression approach and the state-of-the-art hardware fuzzer, DiffuzRTL, respectively.}
Finally, compared to the industry-standard formal verification tool, JasperGold, \ourtool{} does not need human intervention and overcomes its other limitations.

\noindent {\bf Responsible disclosure.}
The bugs have been responsibly disclosed through the legal department of our institution(s).

\section*{Acknowledgement}
Our research work was partially funded by the US Office of Naval Research (ONR Award \#N00014-18-1-2058), by Intel's Scalable Assurance Program, by the Deutsche Forschungsgemeinschaft (DFG, German Research Foundation)—SFB 1119—236615297 within project S2, and by the German Federal Ministry of Education and Research and the Hessian State Ministry for Higher Education, Research and the Arts within ATHENE. We thank Kevin Laeufer from UC Berkeley and 
Jaewon Hur from Seoul National University for helping us with their timely responses related to their respective prior works. 
We thank the TAMU HPRC for their help in providing the software support to generate the data. 
And, we thank anonymous reviewers for their comments.
Any opinions, findings, conclusions, or recommendations expressed herein are those of the authors, and do not necessarily reflect those of the US Government.

\bibliographystyle{ieeetr}
\bibliography{sample}

\appendix
\section*{Appendix}

\section{Mutation Techniques}\label{App:MutationTypes}
\begin{table}[tbh!]
\centering
\caption[Mutation Techniques]{Mutation techniques used by \ourtool{}. }
\resizebox{0.5\textwidth}{!}{%
\begin{tabular}{|l|l|l|}
\hline
\textbf{\#} & \textbf{Name} & \textbf{Description}  \\
\hline
M0 & Bitflip 1/1 & Flip single bit \\
M1 & Bitflip 2/1 & Flip two adjacent bits\\
M2 & Bitflip 4/1 & Flip four adjacent bits \\
M3 & Bitflip 8/8 & Flip single byte \\
M4 & Bitflip 16/8 & Flip two adjacent bytes \\
\hline
M5 & Arith 8/8 & \begin{tabular}{@{}l@{}}
Treat single byte as 8-bit integer,\\ +/- value from 0 to 35 
\end{tabular} \\
M6 & Arith 16/8 & \begin{tabular}{@{}l@{}}
Treat 2 adjacent bytes as 16-bit integer, \\ +/- value from 0 to 35
\end{tabular} \\
M7 & Arith 32/8 & \begin{tabular}{@{}l@{}}
Treat 4 adjacent bytes as 32-bit integer,\\  +/- value from 0 to 35
\end{tabular} \\
\hline
M8 & Random 8 & Overwrite random byte with random value \\
M9 & Delete & Delete an instruction \\
M10 & Clone & Clone an instruction\\
M11 & Opcode & Overwrite opcode bits \\
\hline

\end{tabular}
}
\label{tab:mutations}
\end{table}

Our mutation techniques are inspired by the popular binary manipulation fuzzer, American Fuzzing Loop (AFL)~\cite{citeAflWhite}. 
We use 12 distinct mutation techniques across three categories, as indicated in Table \ref{tab:mutations}.
Techniques M0-M7 mutate only the \textit{data bits} of the instruction.
The mutations M0-M4 perform bitflip operations on one or multiple of the \textit{data bits}. 
The location of the bits is selected randomly from the \textit{data bits}. 
M5-M7 {\it add} or {\it subtract} a random integer, treating one or multiple bytes of \textit{data bits} as a single binary number.
 M8-M12 mutate both the \textit{data bits} and the \textit{opcode bits} and thus could change the instruction type. 
M8 updates a random byte in the instruction with a random value.
M9 replaces the instruction being mutated with a dummy no operation (NOP) instruction. 
M10 clones a random instruction from the test instructions (TIs) and replaces the instruction being mutated with this cloned instruction. 
M11 mutation is targeted specifically to mutate the \textit{opcode bits} of the instruction.
This helps trigger bugs that are related to the control path or \textit{illegal} instructions like the B3 bug.

\section{More Details on the Bugs Detected}\label{App:bugsFound}

We now present more details to ease the understanding of the bugs presented in Section~\ref{sec:bugs}.

\noindent {\bf Bug}~\ref{b2}. 
The expected behavior is to forward the correct exception to the later stages of the processor. 
Exceptions are used to indicate the occurrence of special events like illegal instruction, page faults, or misaligned memory in the processor.
Instruction access fault is one such event that occurs when the program tries to access an instruction memory location to which it does not have permissions. 
This bug occurs because \ariane{} throws a hard-coded exception, \textit{instruction page-fault}, instead of the \textit{instruction access-fault} exception for an instruction access-fault event.

\noindent {\bf Bug}~\ref{b3}
is found in the decode stage, which should reject any instruction that is not listed in the RISC-V specification~\cite{riscv_home} by throwing an \textit{illegal} exception, thereby ensuring no illegal instructions are executed.
This ensures that the processor does not execute any illegal instructions. 
Failure to do so can have severe security implications~\cite{strupe2020uncovering,domas2018hardware}.

\noindent {\bf Bug} ~\ref{b7}.
The Exception Effective Address Register (\texttt{EEAR}) stores the effective address (EA) of the instruction that causes an exception.
EA is used by the exception handling code; e.g., it is used to determine the correct page to load during a page fault exception.
It should be accessible from the supervisor privilege mode according to the OpenRISC specification~\cite{openrisc_home}.

\noindent {\bf Bug}~\ref{b8}.
Pipelined processors are prone to data hazards like the read after write (RAW) hazard, where two consecutive instructions try to write into and read from the same register. 
Hence, the read instruction should wait for the previous write instruction to complete updating the register value, stalling the pipeline.
To improve the performance during hazards, processors use the register forwarding technique, which fetches the value of the register from in-between the pipeline and provides to the read instruction instead of waiting for the previous instruction to complete. 
Register forwarding is a micro-architectural feature and should not affect the architectural state, such as the value of the general purpose registers (GPRs). 
Also, as per the OpenRISC specification~\cite{openrisc_home}, one of the GPRs, the \texttt{GPR0}, should be hard-coded to \textit{0}. 
Hence, this register is used frequently in software programs as a replacement for \textit{0} to optimize the use of this commonly used value. 
Note that the OpenRISC specification~\cite{openrisc_home} expects software to not write into the \texttt{GPR0}. 
This bug would not be detected with traditional approaches which generate inputs following the specification, emphasizing the need for hardware fuzzers like \ourtool{} which verify the hardware even for out-of-spec inputs.

\noindent {\bf Bug}~\ref{b9}.
\orth{} uses multiple flags in the ALU to indicate the events like \textit{carry} and \textit{overflow}. 
Software techniques use this flag to detect integer overflows, whose failure can cause serious security errors or exploitable vulnerabilities like buffer overflow~\cite{dietz2015understanding,wang2010ricb}. 

\begin{lstlisting}[language=C++,belowskip=-15pt,label = {listing:ariane_exploit}, caption={Ariane FENCE.I exploit pseudocode.},style=customcArianeExploit,float=tp,aboveskip=0pt]
char *safe_read(char * data_in) {
  load_canary_into_reg();// set stack canary
  char buf[16]; 
  for (i=0; i<sizeof(buf); ++i) // bounds
    buf[i] = data_in[i];// read data
  check_canary_reg();   // stack canary check
  return buf;
}
void movs() {
  asm ("mov a5, `STACK_OFFSET");
  asm ("mov a5, `STACK_OFFSET");
  ... // continues for several memory pages
  return;
}
void never_called() {
  printf("attack successful\n");
}
int main() {
  jit_load(safe_read);  
  jit_execute();        // safe_read program
  jit_load(movs);       
  asm ("FENCE.I 20\n"); // buggy FENCE.I
  inv_length_chk(); // invalidate bounds check
  inv_canary_chk(); // invalidate canary check
  jit_execute();        // movs program
  return 0;
}
\end{lstlisting}

\section{Exploiting \ariane{} FENCE.I bug}\label{App:ariane_exploit}
This exploit leverages \ref{b1} and \ref{b4} in  \ariane{}. 
Ariane uses a modified Harvard architecture, which is used in x86, ARM, and other processor families.
It contains separate instruction and data caches while allowing the instruction memory also to be accessed as data.
This can lead to instruction and data cache incoherence if the same instruction memory is modified as data and also used for execution. 
These scenarios are common in Just-In-Time (JIT) compilation, where the compiler compiles and writes a program into the instruction memory and runs it during its execution. 
To mitigate cache incoherence, designers either invalidate both caches on write if they share the same memory, introduce instructions to flush, or selectively invalidate the instruction cache.
In the case of RISC-V, \textit{FENCE.I} instruction~\cite{riscv_home} serves this purpose.

\noindent{\bf Exploit.}
The exploit establishes a scenario wherein a software performs basic JIT code generation behavior to load and execute an arbitrary safe JIT program, \textit{safe\_read}, which is widely used nowadays~\cite{google_chrome,bpf,qemu_tcg}.
The \textit{safe\_read} program performs a simple operation of loading data from the location pointed to by its input argument \textit{data\_in} into a local buffer. 
The \textit{safe\_read} includes bounds checking and stack canary safeguards as shown in Line 4 and Line 6, respectively, in Listing~\ref{listing:ariane_exploit}.
To demonstrate the severity of the bugs, the exploit (i)~successfully overcomes both bounds checking and stack canary safeguards, (ii)~constructs a basic return-oriented programming (ROP) chain, and thus, (iii)~executes code unreachable by any valid control flow present in the program.
The exploit code contains a \textit{FENCE.I} instruction every time after the instruction memory is modified, as required by the specification, but one of them is a \textit{failing-FENCE.I} triggering the bug.
We validated that the exploit is not successful on the \textit{spike} RISC-V ISA simulator since it does not have the bug.

\noindent{\bf Exploit flow.} 
Listing~\ref{listing:ariane_exploit} shows the pseudocode for the exploit. 
The programs used by the JIT compiler are pre-compiled. 
The JIT load function, \textit{jit\_load}, loads the \textit{safe\_read} program into the JIT execution space (the space in memory used by the JIT compiler to load and execute programs).
\textit{jit\_execute} executes \textit{safe\_read}, thereby loading the data in the JIT execution space to the instruction cache.
The execution completes safely without stack buffer overflow due to the bounds and canary checks in \textit{safe\_read}.

This attack exploits the \blue{FENCE.I and }cache incoherency bugs, \ref{b1} and \ref{b4},  
by loading a different program \textit{movs} into memory.
The \textit{failing-FENCE.I} is used to retain the cache incoherence in \ariane{} while still following the specification.
\ref{b4} ensures that this failure to fix the cache incoherency is undetected.
Still, this is not enough to cause the stack overflow since calling the \textit{jit\_execute} will execute the \textit{safe\_read} from the instruction cache with its bounds check still intact.  

To prevent this bounds check, the cache line containing the instructions that set the size for the bounds check is invalidated by repeatedly jumping to different addresses contained in that cache line.
Similarly, the stack canary check instructions are also invalidated from the cache to prevent the detection of the stack overflow before returning.
Thus, on executing the \textit{jit\_execute} function, the \textit{safe\_read} will execute from the cache except for the specific lines where it encounters a cache miss, namely: the bounds check and the stack canary check. 
Thus, the instruction for this location will be fetched from the instruction memory where the \textit{safe\_read} is already replaced with the \textit{movs} program.
The \textit{movs} program is built such that the new instruction will set incorrect bounds check value, sufficient to overwrite the stack return address (i.e., \texttt{STACK\_OFFSET}), and the stack canary check no longer executes. 

Having disabled both the checks, the \textit{jit\_execute} results in the execution of \textit{safe\_check} with invalid bounds checks, causing a successful stack buffer overflow. 
The stack is corrupted with the data from the \textit{data\_in} argument.
The exploit loads the array, pointed to by \textit{data\_in}, with the \textit{never\_called} function address. 
Thus, upon exit from the \textit{safe\_read}, the processor will jump to the \textit{never\_called} function even though it is never called, resulting in arbitrary code execution inside the JIT compiler.

\noindent{\bf Ramifications.}
Applications relying on specialized functionality introduced in the \textit{imm}, \textit{rs1}, or \textit{rd} fields, and thus, use \textit{failing-FENCE.I} instructions.
Due to \ref{b1} and \ref{b4}, these applications will have the \textit{failing-FENCE.I} instructions skipped as \textit{illegal} instructions, and hence the processor will not clear the cache.
As shown in the exploit, this can lead to arbitrary code execution in the appropriate contexts.

\section{Exploiting \morkx{} EPCR Register Bug}\label{App:mor1kx_exploit}
The state information of the processor, such as the privilege level, program counter (PC),  processor flags, interrupt enable, and cache enable, is maintained using a set of privilege registers.
Since this state information is security-critical, the ISA defines access permissions based on the processor privilege level for each of these registers, and the hardware implementation enforces those access controls. 

{\bf Bug.}
This exploit leverages the bug~\ref{b6} in \morkx{}. 
The \morkx{} hardware design does not check for the processor privilege level when writing to the exception program counter register (EPCR). This makes the EPCR accessible from any privilege level, whereas the specification of the OpenRISC states that EPCR can only be accessed from the machine privilege level.  

{\bf Exploit.}
The overall flow of the attack is as shown in Listing \ref{listing:mor1kx exploit}. The goal of this exploit is privilege escalation by an attacker with only user privilege access to the system. We run the exploit in a baremetal environment that runs the user application function in user mode. Apart from privilege escalation, this exploit also allows the attacker to run a function of his choice after the privilege escalation. This is a more powerful attack than just privilege escalation since the attacker can now run a program of his choice in privilege mode rather than a random function.

\begin{lstlisting}[language=C++, label = {listing:mor1kx exploit}, caption={\morkx{} exploit pseudocode.},style=customcArianeExploit,float=tp,aboveskip=0pt]
void baremetal_entry() {
    baremetal_initialization(); 
    switch_to_user_mode()
    main(); // run user appln in user mode
}
int main() {
    asm ("csrw epcr, attack_code"); // set the epcr value
    rfe; // trigger exception return
    return 0;
}
attack_code() {
    // code here runs in privilege mode
}
\end{lstlisting}

The exploit is executed in three steps. 
The first step involves setting the exception status register (ESR) to a desired value. ESR is used to store the value of SR when an exception is triggered, and the stored value is restored when the processor finishes the exception handling.
In this step, the attacker waits for the ESR to have a value with privilege mode set to the machine privilege level. 
This is not difficult to perform since an exception in the machine privilege level will set the ESR to the machine privilege level. 
In our case, the value of ESR had the privilege mode set to machine privilege level, at the beginning of the user program, thus eliminating the need to search for such a scenario. 
In the second step, the attacker sets the EPCR with the address to which the processor should jump into.
This access to EPCR is possible due to~\ref{b6}. The final step entails executing a return from exception (RFE) instruction. The RFE instruction causes the processor to think that it should return from an exception and performs the register update. This update sets the SR to the value of the ESR and the PC to address in EPCR. Hence, the SR value now holds machine privilege level, i.e., the processor is running in the privilege mode while executing the attacker's code, and  thus, performing privilege escalation and completely  compromising the security of the processor. 

\bgroup
\createlinenumber{70}{...}
\createlinenumber{71}{72}
\createlinenumber{72}{73}
\createlinenumber{73}{74}
\createlinenumber{74}{75}
\createlinenumber{75}{76}
\createlinenumber{76}{77}
\createlinenumber{77}{78}
\createlinenumber{78}{79}
\createlinenumber{79}{80}
\createlinenumber{80}{81}
\createlinenumber{81}{82}
\createlinenumber{82}{83}
\createlinenumber{83}{...}
\createlinenumber{84}{168}
\createlinenumber{85}{...}
\createlinenumber{86}{204}
\createlinenumber{87}{205}
\createlinenumber{88}{206}
\createlinenumber{89}{207}
\createlinenumber{90}{208}
\createlinenumber{91}{209}
\createlinenumber{92}{210}
\createlinenumber{93}{211}
\createlinenumber{94}{212}
\createlinenumber{95}{213}
\createlinenumber{96}{214}
\createlinenumber{97}{215}
\createlinenumber{98}{216}
\createlinenumber{99}{217}
\createlinenumber{100}{218}
\createlinenumber{101}{219}
\createlinenumber{102}{220}
\createlinenumber{103}{221}
\createlinenumber{104}{222}
\createlinenumber{105}{223}
\begin{lstlisting}[language=Verilog, label = {listing:ExampleDesDiffuzV}, caption={\blue{Verilog code of the hardware design in Figure~\ref{fig:ExampleDesign} instrumented by \difuzzrtl{}.}},style=prettyverilog,float,belowskip=-10pt,aboveskip=0pt,firstnumber=61]
  assign _T = flush | en; // @[cmd3.sc 37:30]
  assign _T_2 = pass == ipass; // @[cmd3.sc 41:20]
  assign sel1 = _T_2 | debug_en; // @[cmd3.sc 41:31]
  assign _T_4 = flush & en; // @[cmd3.sc 46:17]
  assign state_f = _T_4 ? FLUSH : state; // @[cmd3.sc 46:22]
  assign _GEN_1 = {{2'd0}, ~sel1}; // @[cmd3.sc 52:21]
  assign _T_6 = _GEN_1 & state_f; // @[cmd3.sc 52:21]
  assign _GEN_2 = {{2'd0}, sel1}; // @[cmd3.sc 52:40]
  assign _T_7 = _GEN_2 & D_READ; // @[cmd3.sc 52:40]
  ...
  assign cache_controller_cov_read_addr = cache_controller_state;
  assign cache_controller_cov_read_data = cache_controller_cov[cache_controller_cov_read_addr]; // @[Coverage map for cache_controller]
  assign cache_controller_cov_write_data = 1'h1;
  assign cache_controller_cov_write_addr = cache_controller_state;
  assign cache_controller_cov_write_mask = 1'h1;
  assign cache_controller_cov_write_en = 1'h1;
  assign en_shl = en;
  assign en_pad = {1'h0,en_shl};
  assign flush_shl = {flush, 1'h0};
  assign flush_pad = flush_shl;
  assign cache_controller_xor0 = en_pad \xor flush_pad;
  assign io_covSum = cache_controller_covSum;
  ...
  always @(posedge clock) begin
    ...
    if (metaReset) begin
      state <= 3'h0;
    end else begin
      state <= _T_6 | _T_7;
    end
    if (metaReset) begin
      vld <= 1'h0;
    end else begin
      vld <= debug_en | _T;
    end
    cache_controller_state <= cache_controller_xor0;
    if (!(cache_controller_cov_read_data)) begin
      cache_controller_covSum <= cache_controller_covSum + 1'h1;
    end
  end
  always @(posedge clock) begin
    if(cache_controller_cov_write_en & cache_controller_cov_write_mask) begin
      cache_controller_cov[cache_controller_cov_write_addr] <= cache_controller_cov_write_data; // @[Coverage map for cache_controller]
    end
  end
\end{lstlisting}
\egroup
        
\section{Coverage Metrics Of Prior Work}
\label{App:other_cov_metrics}
\blue{We now demonstrate why the coverage metrics of \difuzzrtl{}~\cite{difuzz} and \rfuzz{}~\cite{laeufer2018rfuzz} cannot cover the bugs inserted in Figure~\ref{fig:ExampleDesign}.}

\subsection{\difuzzrtl{}'s coverage metric: control-register coverage}
\difuzzrtl{} uses a coverage metric called control-register coverage.
It defines all the registers that drive the select signals of the muxes as control registers.
All the control registers in each module are concatenated to form a single \textit{module\_state} register, and all the possible values of this \textit{module\_state} registers are defined as coverage points.

When applied to the above example, \difuzzrtl{} should concatenate all the registers that drive the select signals of the two MUXes \cc{1} and \cc{3}. 
Thus,
it concatenates \texttt{flush}, \texttt{en}, \texttt{pass}, \texttt{ipass}, and \texttt{debug\_en} registers.
Since there are five 1-bit registers, there are $2^5=32$ possible values, and \difuzzrtl{} considers each of them as coverage points, creating 32 coverage points.
We now discuss in detail why the control-register coverage metric does not cover the two bugs in Figure~\ref{fig:ExampleDesign}.

\begin{lstlisting}[language=Verilog, label = {listing:ExampleDesDiffuzrep}, caption={\blue{\difuzzrtl{}'s output of the hardware design in Figure~\ref{fig:ExampleDesign}. MUX2 is undetected.}},style=prettyverilog,float,belowskip=-5pt,aboveskip=0pt]
============ Finding Control Registers =========
=============================================
cache_controller
---------------------------------------------
numRegs: 9, numCtrlRegs: 2, numMuxes: 1
=============================================                           
=============================================
============ Instrumenting Coverage ============
=============================================
cache_controller
---------------------------------------------
regStateSize: 2, totBitWidth: 2, numRegs: 2
numOptRegs: 2
=============================================
=============================================
============ Instrumenting metaAssert ==========
=============================================
============ Instrumenting MetaReset ===========
=============================================
cache_controller
---------------------------------------------
=============================================
=============================================

============ Instrumentation Summary ===========   
Total number of registers: 9
Total number of control registers: 2
Total number of muxes: 1
Total number of optimized registers: 2
Total bit width of registers: 15
Total bit width of control registers: 2
Optimized total bit width of control registers: 2
Total bit width of muxes: 1
Total bit width of muxes: 1
=============================================
Total FIRRTL Compile Time: 564.1 ms
\end{lstlisting}

       \begin{itemize}[align=parleft,leftmargin=*]
            \item 
             \blue{\difuzzrtl{} detects only certain implementations of MUXes in the RTL code.           
            When a MUX is implemented differently (e.g., as a combination of NOT, AND, or OR gates), \difuzzrtl{}{} fails to detect the MUX and ignores the corresponding control registers. 
            Therefore, it fails to account for certain control registers driving the select signals of such MUX implementations. Consequently, it does not produce coverage points for these control registers.}
            
            \blue{In the controller example in Listing~\ref{listing:ExampleDesChisel}, 
            the combinational logic \cc{4} generates the select signal \texttt{sel1} of MUX ~\cc{3}.
            \difuzzrtl{} cannot detect this MUX because the corresponding RTL code of this MUX is described using combinational logic (Line 51 of Listing~\ref{listing:ExampleDesChisel}: \texttt{state := ((!sel1\ \&\ state\_f) | (sel1\ \&\ D\_READ)))} instead of control flow constructs (like \texttt{when} block at Lines 45--49 of Listing~\ref{listing:ExampleDesChisel}, thereby failing to detect the bug \textit{b1} in \cc{4}. }

\begin{lstlisting}[language=Verilog, label = listing:ExampleDesVerilog, caption={\blue{Verilog code of the hardware design in Figure~\ref{fig:ExampleDesign}.}},style=prettyverilog,float,belowskip=-5pt,aboveskip=0pt,firstnumber=34]
  wire  sel1 = pass == ipass | debug_en; // @[cmd3.sc 41:31]
  wire [2:0] state_f = flush & en ? FLUSH : state; // @[cmd3.sc 46:22 cmd3.sc 47:17 cmd3.sc 49:17]
  wire  _T_5 = ~sel1; // @[cmd3.sc 52:15]
  wire [2:0] _GEN_1 = {{2'd0}, _T_5}; // @[cmd3.sc 52:21]
  wire [2:0] _T_6 = _GEN_1 & state_f; // @[cmd3.sc 52:21]
  wire [2:0] _GEN_2 = {{2'd0}, sel1}; // @[cmd3.sc 52:40]
  wire [2:0] _T_7 = _GEN_2 & D_READ; // @[cmd3.sc 52:40]
  assign io_state_out = state; // @[cmd3.sc 56:19]
  assign io_vld_out = vld; // @[cmd3.sc 57:17]
  always @(posedge clock) begin
    flush <= io_flush_i; // @[cmd3.sc 28:12]
    en <= io_en_i; // @[cmd3.sc 29:9]
    debug_en <= io_debug_en_i; // @[cmd3.sc 30:15]
    pass <= io_pass_i; // @[cmd3.sc 31:11]
    ipass <= io_ipass_i; // @[cmd3.sc 32:12]
    READ <= io_READ_i; // @[cmd3.sc 33:11]
    FLUSH <= io_FLUSH_i; // @[cmd3.sc 34:12]
    state <= _T_6 | _T_7; // @[cmd3.sc 52:32]
    vld <= debug_en | (flush | en); // @[cmd3.sc 37:21]
  end
\end{lstlisting} 

            \blue{To demonstrate this limitation, we 
            compiled the Chisel code (Listing~\ref{listing:ExampleDesChisel}) of the controller, generated the corresponding FIRRTL code, and ran \difuzzrtl{} on it.
            The instrumented Verilog code and output of \difuzzrtl{} instrumentation are shown in Listing~\ref{listing:ExampleDesDiffuzV} and Listing~\ref{listing:ExampleDesDiffuzrep}, respectively. 
            It can be seen from the Lines 28 and 32 of \difuzzrtl{}’s report (Listing~\ref{listing:ExampleDesDiffuzrep}) that \difuzzrtl{} detected only one MUX and two control registers; Lines 78--82 of the instrumented Verilog code (Listing~\ref{listing:ExampleDesDiffuzV}) show that these control registers are \texttt{flush} and \texttt{en}.
            The control registers (\texttt{pass, ipass, debug\_en}) generating the signal \texttt{sel1} of MUX \cc{3} are not included. Consequently, \difuzzrtl{} does not have any coverage point in \cc{4}, thereby failing to detect \textit{b1}.}

            \item \blue{\difuzzrtl{} focuses only on the control-registers that are generating the select signals of MUXes. 
            Thus, \difuzzrtl{} will not cover any combinational logic that does not drive the select signals of the MUXes.
            In the controller example in Listing~\ref{listing:ExampleDesChisel}, the second bug \textit{b2} is in the combinational logic \cc{6}. \difuzzrtl{} cannot detect this bug since it does not cover the registers, \texttt{flush} and \texttt{en}, generating \texttt{vld} in \cc{6} as these registers are not generating the select signals of any MUXes.}

            \blue{We demonstrate this limitation of \difuzzrtl{} using the same instrumented Verilog code (Listing~\ref{listing:ExampleDesDiffuzV}) and the output of \difuzzrtl{} instrumentation (Listing~\ref{listing:ExampleDesDiffuzrep}) . 
            \difuzzrtl{} only reports the two control registers: \texttt{flush} and \texttt{en} generating the select signal \texttt{sel2} of MUX \cc{1} (Lines 78--82 of the instrumented Verilog code in  Listing~\ref{listing:ExampleDesDiffuzV}).
            However, \difuzzrtl{} does not have any coverage point for the signals in the combinational logic \cc{6}, where the bug resides.
            Combinational logic constitutes a significant portion of the hardware design, and thus these bugs cannot be overlooked as rare corner cases. }

       \end{itemize}

        \bgroup
        \createlinenumber{46}{...}
        \createlinenumber{47}{48}
        \createlinenumber{48}{...}
        \createlinenumber{49}{109}
        \createlinenumber{50}{...}
        \createlinenumber{51}{145}
        \createlinenumber{52}{146}
        \createlinenumber{53}{147}
        \createlinenumber{54}{148}
        \createlinenumber{55}{149}
        \createlinenumber{56}{150}
        \createlinenumber{57}{151}
        \createlinenumber{58}{152}
        \createlinenumber{59}{153}
        \createlinenumber{60}{154}
        \createlinenumber{61}{155}
\begin{lstlisting}[language=Verilog, label = {listing:ExampleDesRfuzzV}, caption={\blue{Verilog code of the hardware design in Figure~\ref{fig:ExampleDesign} instrumented by \rfuzz{}.}},style=prettyverilog,float,belowskip=-5pt,aboveskip=0pt,firstnumber=37]
  wire  _T = flush | en; // @[cmd3.sc 37:30]
  wire  _T_2 = pass == ipass; // @[cmd3.sc 41:20]
  wire  sel1 = _T_2 | debug_en; // @[cmd3.sc 41:31]
  wire [2:0] state_f = profilePin ? FLUSH : state; // @[cmd3.sc 46:22]
  wire  _T_5 = ~sel1; // @[cmd3.sc 52:15]
  wire [2:0] _GEN_1 = {{2'd0}, _T_5}; // @[cmd3.sc 52:21]
  wire [2:0] _T_6 = _GEN_1 & state_f; // @[cmd3.sc 52:21]
  wire [2:0] _GEN_2 = {{2'd0}, sel1}; // @[cmd3.sc 52:40]
  wire [2:0] _T_7 = _GEN_2 & D_READ; // @[cmd3.sc 52:40]
  ...
  assign auto_cover_out = flush & en;
  ...
  always @(posedge clock) begin
    ...
    if (metaReset) begin
      state <= 3'h0;
    end else begin
      state <= _T_6 | _T_7;
    end
    if (metaReset) begin
      vld <= 1'h0;
    end else begin
      vld <= debug_en | _T;
    end
  end
\end{lstlisting}
        \egroup

\subsection{\blue{\rfuzz{}'s coverage metric: Mux-coverage}}

       \blue{ \rfuzz{} uses a coverage metric called mux-coverage. It treats the select signal of each 2:1 MUX as a coverage point. When applied to the controller design in Figure~\ref{fig:ExampleDesign}, \texttt{sel1} and \texttt{sel2} signals are selected as the mux-coverage points. Since both are 1-bit wide, the total number of mux-coverage points is $2^1 + 2^1 = 4$ coverage points. We now discuss in detail why the mux-coverage metric does not cover the two bugs in Figure~\ref{fig:ExampleDesign}.}

        \begin{enumerate}
            \item \blue{\rfuzz{} detects only certain implementations of MUXes in the RTL code. When a MUX is implemented differently (e.g., as a combination of NOT, AND, or OR gates), \rfuzz{} fails to detect the MUX and ignores the corresponding select signals. Therefore, it fails to account for select signals of such MUX implementations. Consequently, it does not produce coverage point for these MUXes.}
            
            \blue{In the controller example in Listing~\ref{listing:ExampleDesChisel}, 
            the combinational logic \cc{4} generates the select signal \texttt{sel1} of MUX ~\cc{3}.
            \rfuzz{} cannot detect this MUX because the corresponding RTL code of this MUX is described using combinational logic (Line 51 of Listing~\ref{listing:ExampleDesChisel}: \texttt{state := ((!sel1\ \&\ state\_f) | (sel1\ \&\ D\_READ)))} instead of control flow constructs (like \texttt{when} block at Lines 45--49 of Listing~\ref{listing:ExampleDesChisel}, thereby failing to detect the bug \textit{b1} in \cc{4}. }
            
            
             \blue{To demonstrate this limitation, we 
            compiled the Chisel code (Listing~\ref{listing:ExampleDesChisel}) of the controller, generated the corresponding FIRRTL code, and ran \rfuzz{} on it.
            The instrumented Verilog code and output of \rfuzz{} instrumentation are shown in Listing~\ref{listing:ExampleDesRfuzzV} and Listing~\ref{listing:ExampleDesRfuzzRep}, respectively. 
            It can be seen from the Lines 49 and 58 of \rfuzz{}’s report (Listing~\ref{listing:ExampleDesRfuzzRep}) that \rfuzz{} detected only one select signal of the MUX \cc{1}; Line 47 of the instrumented Verilog code (Listing~\ref{listing:ExampleDesRfuzzV}) shows the same.
            The select signal \texttt{sel1} of MUX \cc{3} is not included. Consequently, \rfuzz{} does not have any coverage point in \cc{4}, thereby failing to detect \textit{b1}. }

            \item \blue{\rfuzz{} focuses only on the select signals of the MUXes. 
            Thus, \rfuzz{} will not cover any combinational logic that does not drive the select signals of the MUXes.
            In the controller example in Listing~\ref{listing:ExampleDesChisel}, the second bug \textit{b2} is in the combinational logic \cc{6}. \rfuzz{} cannot detect this bug since it does not cover the registers, \texttt{flush} and \texttt{en}, generating \texttt{vld} in \cc{6} as these registers are not the select signals of any MUXes.}
            
           \blue{ We demonstrate this limitation of \rfuzz{} using the same instrumented Verilog code (Listing~\ref{listing:ExampleDesRfuzzV}) and the output of \rfuzz{} instrumentation (Listing~\ref{listing:ExampleDesRfuzzRep}) . 
            \rfuzz{} only reports the one signal: the select signal \texttt{sel2} of the MUX \cc{1} (Line 58 of the \rfuzz{}'s output). 
            However, \rfuzz{} does not have any coverage point for the signals in the combinational logic \cc{6}, where the bug resides.
            Combinational logic constitutes a significant portion of the hardware design, and thus these bugs cannot be overlooked as rare corner cases. }

        \end{enumerate}
   
\begin{lstlisting}[language=Verilog, label = {listing:ExampleDesRfuzzRep}, caption={\blue{\rfuzz{}'s output for the hardware design in Figure~\ref{fig:ExampleDesign}. MUX2 is undetected.}},style=prettyverilog,float,belowskip=-5pt,aboveskip=0pt,firstnumber=47]
[[port]]
name = "auto_cover_out"
width = 1

[[coverage]]
port = "auto_cover_out"
name = "_GEN_0"
index = 0
filename = ""
line = -1
column = -1
human = "(flush and en)"
\end{lstlisting}

\end{document}